\documentclass[aps,prd,showpacs,showkeys,preprintnumbers]{revtex4}
\usepackage{graphicx}                              
\usepackage{amsmath,amsfonts}
\graphicspath{{./pictures/}}                       

\usepackage{epsfig}
\usepackage{amssymb}
\usepackage{amsfonts}
\usepackage{float}
\usepackage{bbm}
\usepackage{psfrag}
\newcommand{\nin}{\noindent}

\newcommand{\bdm}{\begin{displaymath}}
\newcommand{\edm}{\end{displaymath}}
\newcommand{\beq}{\begin{equation}}
\newcommand{\eeq}{\end{equation}}
\newcommand{\bea}{\begin{eqnarray}}
\newcommand{\eea}{\end{eqnarray}}
\newcommand{\bit}{\begin{itemize}}
\newcommand{\eit}{\end{itemize}}
\newcommand{\bc}{\begin{center}}
\newcommand{\ec}{\end{center}}
\newcommand{\re}{\relax{\rm I\kern-.18em R}}
\newcommand{\ID}{\mathbbm{1}}

\newcommand{\N}{\mathbbm{N}}

\newcommand{\Comp}{\mathbbm{C}}

\newcommand{\ie}{{\it i.e. }}
\newcommand{\D}{{\cal D}^{(ov)}}
\newcommand{\Dprime}{{\cal D'}^{(ov)}}
\newcommand{\DnoK}{{\breve{\cal D}}^{(ov)}}
\newcommand{\SD}{\hat{\cal D}^{(ov)}}

\newcommand{\SDw}{\hat{\cal D}^{(W)}}
\newcommand{\DDmtwoRhoInv}{{\cal A}}
\newcommand{\ImpSpace}{{\cal P}}
\newcommand{\ImpBasis}{{\cal Q}}
\newcommand{\sumFL}{\sum\limits_{i=1}^{N_f}}

\newcommand{\LagrangeMul}{\lambda}

\begin{document}
\preprint{HU-EP-07/16, DESY 07-063}

\title{The phase structure of a chirally invariant lattice Higgs-Yukawa model\\
for small and for large values of the Yukawa coupling constant} 

\author{P. Gerhold$^a$, K. Jansen$^b$}
\affiliation{$^a$Humboldt-Universit\"at zu Berlin, Institut f\"ur Physik, 
Newtonstr. 15, D-12489 Berlin, Germany\\
$^b$DESY,\\
 Platanenallee 6, D-15738 Zeuthen, Germany}

\date{June 26, 2007}

\begin{abstract}
We consider a chirally invariant lattice Higgs-Yukawa model based on the Neuberger 
overlap operator $\D$. As a first step towards the eventual determination of Higgs 
mass bounds we study the phase diagram of the model analytically 
in the large $N_f$-limit. We present an expression for the effective potential 
at tree-level in the regime of small Yukawa and quartic coupling constants and determine 
the order of the phase transitions. In the case of strong Yukawa couplings the model 
effectively becomes an $O(4)$-symmetric non-linear $\sigma$-model for all values of the
quartic coupling constant. 
This leads to the existence of a symmetric phase also in the regime of large values of 
the Yukawa coupling constant. On finite and small lattices, however, strong finite volume 
effects prevent the expectation value of the Higgs field from vanishing thus obscuring the 
existence of the symmetric phase at strong Yukawa couplings.
\end{abstract}



\maketitle

\section{Introduction}
\label{sec:Introduction}

Non-perturbative investigations of lattice regularized Higgs-Yukawa models  
as a limit of the electroweak sector of the Standard Model have been subject 
of many investigations in the early 1990's, see e.g. the review articles of 
Refs.~\cite{Smit:1989tz,Shigemitsu:1991tc,Golterman:1990nx,book:Montvay,book:Jersak,Golterman:1992ye,
Jansen:1994ym}.
These lattice studies were motivated by the interest in a better understanding of the fermion mass 
generation via the Higgs mechanism on a non-perturbative level. In addition, the focus 
has been on the determination of bounds on the Higgs mass and the Yukawa couplings 
which translate directly into bounds on the - at that time not yet discovered - top quark mass. 
However, these investigations were blocked, since the influence of unwanted fermion doublers could 
not successfully be suppressed. Moreover, the lattice models of these studies suffered from the lack of 
chiral symmetry. The latter, however, would be indispensable for a consistent lattice regularization 
of chiral gauge theories such as the Standard Model of electroweak interactions. 

Here, we want to extend these earlier investigations in a new direction in order to overcome 
the previously encountered drawbacks by following the proposition of L\"uscher~\cite{Luscher:1998pq} 
for a chirally invariant lattice Higgs-Yukawa model based on the Neuberger overlap operator~\cite{Neuberger:1998wv}. 
Within this model an exact lattice chiral symmetry can be established while suppressing the fermion 
doublers at the same time. This is possible despite of the Nielsen-Ninomiya theorem~\cite{Nielsen:1980rz}, 
since the established lattice chiral symmetry is not the continuum chiral symmetry itself, but 
recovers the latter symmetry only in the continuum limit. We consider here a Higgs-Yukawa model including 
only the two heaviest fermions, \ie the top-bottom doublet, and the Higgs fields. This simplification 
is reasonable, since the fermion-Higgs coupling is proportional to the fermion mass and hence small 
for the light doublets. We also neglect any gauge fields within this model, since they can be taken into 
account via perturbation theory.

As a first step towards a numerical investigation of this Higgs-Yukawa model we begin by studying
its phase structure. Here we present an analytical investigation of the phase diagram in the large 
$N_f$-limit following Refs.~\cite{Hasenfratz:1991it,Hasenfratz:1992xs}. We refer the reader to these 
references for earlier works on lattice Higgs-Yukawa models. (See also Ref.~\cite{Gerhold:2006rc} for
a first account of our work.) In the present paper we access the phase structure of the model at small
and at large values of the Yukawa coupling constant, putting particular emphasis on the existence
of a symmetric phase also in the strong Yukawa coupling regime. The latter strong coupling regime of
a closely related, chirally invariant Higgs-Yukawa model in two dimensions was also studied in the recent 
work~\cite{Giedt:2007qg} and corresponding Monte-Carlo simulations, performed in that model, support 
the existence of such phase~\cite{Giedt:2007pri}.
Extensions of our present paper, in particular concerning the verification
of the analytically obtained phase structure by explicit numerical simulations and addressing the 
question of lower and upper bounds on the Higgs boson mass, will be discussed in forthcoming publications.

The outline of this paper is as follows:
In Section~\ref{chap:model} we briefly describe the Higgs-Yukawa model considered here. In the following 
Section~\ref{sec:SmallYukawaCouplings} we derive an expression for the effective potential in terms of the
amplitudes of the constant and staggered modes of the Higgs field, which is a reasonable approximation 
at small values of the Yukawa and quartic coupling constants. We then present the resulting phase diagram 
in the large $N_f$-limit and determine the order of the occurring phase transitions. The phase 
structure in the regime of large values of the Yukawa coupling constant and arbitrary quartic coupling
constants is then accessed by means of a different large $N_f$-limit presented in Section~\ref{sec:LargeYukawaCouplings}. 
We show that a symmetric phase also exists at large Yukawa coupling constants. On small lattices, 
however, this symmetric phase is shadowed by finite volume effects preventing the expectation value of 
the Higgs field from vanishing. We then end with a short summary and outlook.

\section{The model}
\label{chap:model}
Aspiring to investigate the Higgs Sector of the Standard Model of electroweak interactions, 
we consider here a four-dimensional, chirally invariant lattice Higgs-Yukawa model 
containing one four-component, real Higgs field $\Phi$ and a number of $N_f$ fermion doublets.
The latter are represented by eight-component spinors $\psi^{(i)}$, $\bar\psi^{(i)}$ with 
$i=1,...,N_f$. However, these $N_f$ doublets are all degenerated within this model and correspond 
to the heaviest fermion doublet only, \ie to the top-bottom doublet. This is a reasonable 
simplification due to the fermion-Higgs coupling being proportional to the fermion mass. 
We have introduced the fermion doublet number $N_f$ nonetheless, because it will be possible 
to access the model analytically in the limit of large numbers of (degenerated) fermion doublets. 

Furthermore, there are also $N_f$ {\it auxiliary} fermionic doublets $\chi^{(i)}$, $\bar\chi^{(i)}$ 
present in the model, which serve as a construction tool for the creation of a chirally 
invariant Yukawa interaction term. However, once the chiral invariance is established, these 
{\it unphysical} fields can be integrated out leading to a more complicated model depending then 
only on the Higgs field $\Phi$ and the $N_f$ {\it physical} fermion doublets $\psi^{(i)}$. 
The partition function of the given model can be written as 
\beq
Z = \int D\Phi\,\prod\limits_{i=1}^{N_f} \left[D\psi^{(i)}\, D\bar\psi^{(i)}\, D\chi^{(i)}\,
D\bar\chi^{(i)} \right]\,  \exp\left( -S_\Phi -S_F^{kin} - S_Y  \right)
\eeq
with the total action being decomposed into the Higgs action $S_\Phi$, the kinetic fermion action $S_F^{kin}$, 
and the Yukawa coupling term $S_Y$. It should be stressed once again that {\it no gauge fields} 
are included within this model. 

The four-dimensional space-time lattice, that the model is discretized upon, is assumed to have  
$L$ lattice sites per dimension such that its total volume is $V=L^4$. Here we allow for both, 
finite size lattices with even $L\in \N $ as well as lattices with infinite extension, \ie $L=\infty$,
and we set the lattice spacing $a$ to unity for convenience. The kinetic fermion 
action describing the propagation of the physical fermion fields $\psi^{(i)}$,$\bar\psi^{(i)}$ 
is then given in the usual manner according to
\beq
S_F^{kin} = \sumFL \sum\limits_{n,m} \bar\psi^{(i)}_n \D_{n,m} \psi^{(i)}_{m} -
2\rho\bar\chi^{(i)}_n\ID_{n,m} \chi^{(i)}_m
\eeq
where the four-dimensional coordinates $n,m$ as well as all field variables and coupling constants 
are given in dimensionless lattice units throughout this paper.  Here, the (doublet) Dirac operator 
$\D= \SD\otimes\SD$ is given by the Neuberger overlap operator $\SD$, which is related to the Wilson 
operator $\SDw=\gamma^E_\mu \frac{1}{2}(\nabla_\mu^f + \nabla_\mu^b) - \frac{r}{2} \nabla^b_\mu\nabla^f_\mu$ 
by 
\bea
\SD &=& \rho\left\{1+\frac{\hat A}{\sqrt{\hat A^\dagger \hat A}}   \right\},\quad \hat A = \SDw - \rho, \quad 1\le \rho < 2r
\eea
with $\nabla^f_\mu$, $\nabla^b_\mu$ denoting the forward and backward difference quotients, respectively. 
In absence of gauge fields the eigenvectors and eigenvalues of the Neuberger operator are explicitly known. 
In momentum space with the allowed four-component momenta
\beq
p \in \ImpSpace = \Bigg\{
\begin{array}{*{3}{c}}
(-\pi,\pi]^{\otimes 4} & : & \mbox{for } L = \infty \\
\{2\pi n/L\,:\, n\in \N_0, n<L\}^{\otimes 4} & : & \mbox{for } L < \infty \\
\end{array}
\eeq
the eigenvectors of the doublet operator $\D$ are given as
\beq
\Psi_n^{p,\zeta\epsilon k} = e^{i p \cdot n} \cdot u^{\zeta\epsilon k}(p),\quad
u^{\zeta\epsilon k}(p) = \sqrt{\frac{1}{2}}
\left(
\begin{array}{*{1}{c}}
u^{\epsilon k}(p) \\
\zeta u^{\epsilon k}(p) \\
\end{array}
\right), \quad
\zeta=\pm 1, \,
\epsilon=\pm 1, \,
k \in \{1,2\}
\eeq
with $u^{\epsilon k}(p)$ denoting the usual four-component spinor structure 
\beq
u^{\epsilon k}(p) = 
\sqrt{\frac{1}{2}}
\left(
\begin{array}{*{1}{c}}
\xi_k \\
\epsilon\frac{\tilde p \bar\Theta}{\sqrt{\tilde p^2}} \xi_k 
\end{array}
\right)
\,
\mbox{for }\tilde p\neq 0
\quad
\mbox{and}
\quad
u^{\epsilon k}(p) = 
\sqrt{\frac{1}{2}}
\left(
\begin{array}{*{1}{c}}
\xi_k \\
\epsilon \xi_k \\
\end{array}
\right)
\,
\mbox{for }\tilde p= 0,
\eeq
respectively.
Here $\xi_k\in\Comp^2$ are two orthonormal vectors and the four component 
quaternionic vectors $\Theta$, $\bar\Theta$
are defined as $\Theta = (\ID, -i\vec\tau)$ and 
$\bar\Theta = (\ID, +i\vec\tau) = \Theta^\dagger$ with $\vec\tau$
denoting the vector of Pauli matrices. It is well known that the eigenvalues $\nu^\pm(p)$ of 
$\SD$ with $\mbox{Im}[\nu^\pm(p)] \gtrless0$ 
form a circle in the complex plane, the radius of which is given by the parameter $\rho$. 
These eigenvalues are explicitly given 
by
\bea
\nu^\epsilon(p)&=& \rho + \rho\cdot\frac{\epsilon i\sqrt{\tilde p^2} + 2r\hat p^2 - \rho}{\sqrt{\tilde p^2 + (2r\hat p^2 -
\rho)^2}},\quad \tilde p_\mu = \sin(p_\mu),\quad \hat p_\mu = \sin\left(\frac{p_\mu}{2}\right).
\eea
The auxiliary fields $\chi^{(i)}$ on the other hand do not propagate at all and their 
contribution to $S_F^{kin}$
is chosen such that the model will obey an exact  lattice chiral symmetry.

The Higgs field couples to the fermions according to the Yukawa coupling term
\beq
S_Y = y_N \sum\limits_{n,m}\sumFL(\bar\psi^{(i)}_n+\bar\chi^{(i)}_n) \underbrace{\left[
\ID_{n,m}\frac{(1-\gamma_5)}{2}\phi_n 
+ \ID_{n,m}\frac{(1+\gamma_5)}{2}\phi^{\dagger}_n  \right]}_{B_{n,m}} (\psi^{(i)}_m+\chi^{(i)}_m)
\label{eq:DefYukawaCouplingTerm}
\eeq
where $y_N$ denotes the Yukawa coupling constant and $B_{n,m}$ will be referred to as Yukawa coupling matrix. 
Here the Higgs field $\Phi_n$ is rewritten as a quaternionic, $2 \times 2$ matrix 
$\phi_n = \Phi_n^0\ID -i\Phi_n^j\tau_j$ acting on the flavor index of the fermionic
doublets. Due to the chiral character of this model, left- and right-handed fermions couple differently to the 
Higgs field, as can be seen from the appearance of the projectors $(1\pm \gamma_5)/2$ in the Yukawa term.
Multiplying out the involved Gamma- and Pauli-matrices one can rewrite the Yukawa coupling matrix in the 
compactified form
\bea
B_{m,n} = \delta_{m,n} \cdot \hat B(\Phi_n), \quad
\hat B(\Phi_n) =  
\left(
\begin{array}{*{2}{c}}
\Phi_n^0\ID+i\Phi_n^3\gamma_5 & \Phi_n^2\gamma_5 +i\Phi_n^1\gamma_5  \\
-\Phi_n^2\gamma_5 +i\Phi_n^1\gamma_5 & \Phi_n^0\ID-i\Phi_n^3\gamma_5  \\
\end{array}
\right)
\label{eq:DefBHat}
\eea
being block diagonal in position space.
The model then obeys an exact, but lattice modified, chiral symmetry according to
\bea
\delta\psi^{(i)}=i\epsilon\left[\gamma_5\left(1-\frac{1}{2\rho}\D \right)\psi^{(i)} + \gamma_5\chi^{(i)}\right], & 
\delta\chi^{(i)} = i\epsilon\gamma_5\frac{1}{2\rho}\D\psi^{(i)},& 
\delta\phi =2i\epsilon\phi    \\
\delta\bar\psi^{(i)}=i\epsilon\left[\bar\psi^{(i)}\left(1-\frac{1}{2\rho}\D \right)\gamma_5 + \bar\chi^{(i)}\gamma_5\right], & 
\delta\bar\chi^{(i)} = i\epsilon\bar\psi^{(i)}\frac{1}{2\rho}\D\gamma_5,& 
\delta\phi^\dagger =-2i\epsilon\phi^\dagger    
\eea
with $\epsilon$ denoting here the infinitesimal parameter of the chiral transformation.
Since the (here omitted) lattice spacing $a$ appears in front of the Dirac operators,
this exact symmetry recovers the continuum chiral symmetry (after having integrated out the 
auxiliary fermion fields) in the continuum limit~\cite{Luscher:1998pq}.

Finally, we use a slightly unusual notation for the Higgs action $S_\Phi$ given by
\beq
\label{eq:ModPhiAction}
S_\Phi = -\kappa_N\sum_{n,\mu} \Phi_n^{\dagger} \left[\Phi_{n+\hat\mu} + \Phi_{n-\hat\mu}\right]
+ \sum_{n} \Phi^{\dagger}_n\Phi_n + \lambda_N \sum_{n} \left(\Phi^{\dagger}_n\Phi_n - N_f \right)^2
\eeq
where $\kappa_N$ denotes the hopping parameter and $\lambda_N$ is the quartic coupling constant. This notation with
$N_f$ appearing in the quartic coupling term (which turns out to be more convenient for the later
analytical considerations) can easily be shown to be equivalent to the more commonly used lattice version
of the $\varphi^4$-action
\bea
\label{eq:UsualPhiAction2}
S_{\varphi} &=&  -\kappa\sum_{n,\mu} \varphi_n^{\dagger} \left[\varphi_{n+\hat\mu} + 
\varphi_{n-\hat\mu}\right]+ \sum_{n} \varphi^{\dagger}_n\varphi_n + 
\lambda \sum_{n} \left(\varphi^{\dagger}_n\varphi_n - 1 \right)^2
\eea
by rescaling the coupling constants $\lambda, \kappa, y$ and the Higgs field $\varphi$ according to
\beq
\Phi =  C\cdot\varphi,\quad \lambda_N = \frac{\lambda}{ C^4},\quad \kappa_N = \frac{\kappa}{ C^2},\quad y_N =\frac{y}{ C},
\eeq
where the constant $C$ has to obey the condition
\beq
\label{eq:DetOfScaleFacC}
C^2 - 2\lambda_NN_f C^2 = 1-2\lambda.
\eeq
Furthermore, this latter action in Eq.~(\ref{eq:UsualPhiAction2}) can also easily be connected to 
the usual continuum notation 
\bea
\label{eq:UsualPhiAction}
S_{\hat \varphi} &=& \sum_n \left\{\frac{1}{2}\left(\nabla^f_\mu\hat\varphi\right)_n^{\dagger} \nabla^f_\mu\hat\varphi_n 
+ \frac{1}{2}m_0^2\hat\varphi_n^{\dagger}\hat\varphi_n + \lambda_0\left(\hat\varphi_n^{\dagger}\hat\varphi_n\right)^2   \right\},
\eea
which explicitly involves a bare mass $m_0$ and the forward difference quotient $\nabla^f_\mu$. This 
connection is established by scaling the field and coupling constants according to 
\beq
\varphi_n = \frac{\hat\varphi_n}{\hat C},\quad \lambda = \hat C^4\cdot\lambda_0,\quad 
\kappa = \frac{\hat C^2}{2},\quad y = y_0 \cdot \hat C
\eeq
where the constant $\hat C$ has to obey the relation
\beq
1 = \hat C^2\cdot \left(\frac{m_0^2+8}{2}+2\lambda_0 \hat C^2\right).
\eeq

For the further analytical treatment of this model we integrate out the
fermionic degrees of freedom leading to an effective Higgs model given by
\beq
\label{eq:DefEffectiveHiggsModelPartitionF}
Z = \int D\Phi\, \exp\left( -S_{eff}[\Phi] \right)
\eeq
with the effective action $S_{eff}[\Phi]$ defined as
\beq
\exp\left( -S_{eff}[\Phi] \right) = \int \prod\limits_{i=1}^{N_f}
\left[D\psi^{(i)} D\bar\psi^{(i)} D\chi^{(i)} D\bar\chi^{(i)} \right]
\exp\left(-S_\Phi -S_F^{kin} - S_Y  \right).
\label{eq:DefEffectiveAction}
\eeq
By applying some adequate substitutions the Grassmann integrations can be
performed allowing to write the effective action $S_{eff}[\Phi]$ in terms
of fermionic determinants according to
\beq
\label{eq:effectiveHiggsAction1}
S_{eff}[\Phi]= S_\Phi[\Phi] - N_f\cdot \log\left[\det\left( y_NB\D -2\rho\D -2\rho y_N B \right) \right].
\eeq

\section{Large $N_f$-limit for small Yukawa coupling parameters}
\label{sec:SmallYukawaCouplings}
In this section we will derive the phase structure of the introduced
Higgs-Yukawa model in the large $N_f$-limit for small values of the 
Yukawa and quartic coupling constants. The idea is to factorize the number 
of involved fermion doublets $N_f$ out of the effective action $S_{eff}[\Phi]$,
since the integral over all Higgs field configurations in Eq.~(\ref{eq:DefEffectiveHiggsModelPartitionF})
can then be reduced to the sum over all absolute minima of the effective 
action when sending $N_f$ to infinity. This factorization can be achieved 
by scaling the coupling constants and the Higgs field itself according to 
\beq
\label{eq:LargeNBehaviourOfCouplings1}
y_N = \frac{\tilde y_N}{\sqrt{N_f}}\,,\quad  
\lambda_N = \frac{\tilde \lambda_N}{N_f}\,,\quad
\kappa_N = \tilde \kappa_N\,,\quad
\Phi_n = \sqrt{N_f} \cdot\tilde\Phi_n\,,
\eeq
where the quantities $\tilde y_N$, $\tilde \lambda_N$, $\tilde \kappa_N$,
and $\tilde\Phi_n$ are kept constant in the limit $N_f\rightarrow\infty$.

One is thus left with the problem of finding the absolute minima of 
$S_{eff}[\Phi]$ in terms of the latter quantities. In general the operators
$B$ and $\D$ do not share a common eigenvector basis making the analytical
evaluation of the determinant in Eq.~(\ref{eq:effectiveHiggsAction1})
impossible for general, space-time dependent Higgs fields.
However, for sufficiently small values of the Yukawa and quartic coupling constants 
the kinetic term of the Higgs action becomes dominant allowing to restrict 
the search for the absolute minima of $S_{eff}[\Phi]$ to the ansatz
\beq
\label{eq:staggeredAnsatz}
\Phi_n = \hat\Phi \cdot \sqrt{N_f} \cdot \left(m + s\cdot (-1)^{\sum\limits_{\mu}n_\mu}    \right) 
\eeq
taking only a constant and a staggered mode of the Higgs field into account.
Here $\hat\Phi\in \re^4$ denotes a constant 4-dimensional unit vector 
($|\hat\Phi|=1$), and we will refer to $m,\, s\in\re$ in the following as 
magnetization and staggered magnetization, respectively. 

For the actual evaluation of the effective action we use the fact that 
the matrix $B$ now has a diagonal-plus-subdiagonal-block-structure in 
momentum space due to the chosen ansatz for the Higgs field according to 
\bea
\label{eq:StructureOfB}
\left[\D - \frac{y_N}{2\rho}B \left(\D-2\rho\right) \right](p_1,p_2) &=&
- \frac{\tilde y_N}{2\rho}\Bigg[ m\cdot\delta(p_1,p_2)\cdot\hat B^{(p_2)}(\hat\Phi) \cdot \left(\D(p_2) - 2\rho \right)\\ 
&+&s\cdot\delta(p_1,\wp_2)\cdot U(p_1,p_2) \cdot \hat B^{(p_2)}  \cdot \left(\D(p_2) -2\rho \right)  \Bigg] \nonumber\\
&+&\delta(p_1,p_2)\cdot\D(p_2) \nonumber \,,
\eea
where the diagonal part is caused by the constant mode of the Higgs field, while the sub-diagonal
contribution is created by the staggered mode. In Eq.~(\ref{eq:StructureOfB}) this is expressed
by $\wp_2$ denoting the shifted momenta $\wp_2 = p_2 + (\pi,\pi,\pi,\pi)$, where
adequate modulo-operations are implicit to guarantee that $\wp_2\in\ImpSpace$. The 
matrices $U(p_1,p_2)$, $\D(p)$, and $\hat B^{(p)}$ are $8\times 8$-matrices with the indices 
$\zeta_1\epsilon_1 k_1,\zeta_2\epsilon_2 k_2$ and
denote the spinor basis transformation matrix
\beq
\label{eq:DefOfSpinorBasisTransMat}
U(p_1,p_2)_{\zeta_1\epsilon_1 k_1,\zeta_2\epsilon_2 k_2} = 
\left[u^{\zeta_1\epsilon_1 k_1}(p_1)\right]^\dagger u^{\zeta_2\epsilon_2 k_2}(p_2),
\eeq
the Dirac matrix
\beq
\label{eq:MatDiracSpinorRep}
\D(p)_{\zeta_1\epsilon_1 k_1,\zeta_2\epsilon_2 k_2} = \delta_{\epsilon_1,\epsilon_2}\cdot\delta_{k_1,k_2}\cdot
\delta_{\zeta_1,\zeta_2} \cdot \nu^{\epsilon_1}(p),
\eeq
and the Yukawa coupling matrix

\bea
\label{eq:MatBinSpinorRep}
\hat B^{(p)}(\hat\Phi)_{\zeta_1\epsilon_1k_1,\zeta_2\epsilon_2k_2}&=&\left[u^{\zeta_1\epsilon_1k_1}(p)\right]^\dagger 
\hat B(\hat\Phi)\, u^{\zeta_2\epsilon_2k_2}(p) \ \\
&=& \delta_{k_1,k_2} \Big[\delta_{\epsilon_1,\epsilon_2}\delta_{\zeta_1,\zeta_2}\cdot \hat\Phi^0
+\delta_{\epsilon_1,-\epsilon_2}\Big\{i\zeta_2\delta_{\zeta_1,\zeta_2}\hat\Phi^1 +
\delta_{\zeta_1,-\zeta_2}\left[i\hat\Phi^3+\zeta_2\hat\Phi^2\right]\Big\}\Big], \nonumber
\eea

\nin
respectively. Due to this diagonal-subdiagonal-block-structure the determinant 
in Eq.~(\ref{eq:effectiveHiggsAction1}) can thus be factorized by merging 
the four $8\times 8$ blocks, which correspond to the momentum indices ${(p,p)}$, ${(\wp,p)}$, ${(p,\wp)}$, 
and ${(\wp,\wp)}$. Up to some constant terms, which are independent of $\Phi$, we can thus rewrite the 
effective action as
\bea
S_{eff}[\Phi]&=& S_\Phi[\Phi] - N_f\cdot \log\left[\det\left(\D - \frac{y_N}{2\rho}\cdot B\cdot \left(\D-2\rho\right) \right) \right]\\
&=& S_\Phi[\Phi] - N_f\cdot \log\Bigg[\prod\limits_{{p\in\ImpSpace}\atop{0\le p_3<\pi}} {\det \left(\D(p)\otimes\D(\wp)-\frac{\tilde y_N}{2\rho}{\cal M}(p)\right)} 
\Bigg],
\label{eq:EffActionRewrittenForSmallY}
\eea
where the restriction $0\le p_3<\pi$ has just been introduced to prevent the double counting that would
occur if one would have performed the product over all $p\in\ImpSpace$ after having merged the blocks.
Here ${\cal M}(p)$ denotes these merged, momentum dependent $16 \times 16$ matrices given by
\beq
{\cal M}(p) = 
\left(
\begin{array}{*{2}{c}}
{\cal M}^{1,1}(p) &{\cal M}^{1,2}(p) \\
{\cal M}^{2,1}(p) &{\cal M}^{2,2}(p) \\
\end{array}
\right)
\eeq
with
\bea
{\cal M}^{1,1}(p) &= & m\cdot \hat B^{(p)}(\hat\Phi) \cdot \left(\D(p)-2\rho\right),  \\
{\cal M}^{1,2}(p) &= & s\cdot U(p,\wp)\cdot \hat B^{(\wp)}(\hat\Phi) \cdot \left(\D(\wp)-2\rho\right),  \\
{\cal M}^{2,1}(p) &= & s\cdot U(\wp,p)\cdot \hat B^{(p)}(\hat\Phi) \cdot \left(\D(p)-2\rho\right),  \\
{\cal M}^{2,2}(p) &= & m\cdot \hat B^{(\wp)}(\hat\Phi) \cdot \left(\D(\wp)-2\rho\right). 
\eea
The expression in Eq.~(\ref{eq:EffActionRewrittenForSmallY}) can be written 
more compactly, taking the fact into account that the matrices involved in that
expression are diagonal with respect to the index $k$ due to
Eq.~(\ref{eq:MatDiracSpinorRep}), Eq.~(\ref{eq:MatBinSpinorRep}) and
\beq
U(p,\wp)_{\zeta_1\epsilon_1 k_1,\zeta_2\epsilon_2 k_2} = \delta_{\zeta_1,\zeta_2} \cdot
\delta_{\epsilon_1,-\epsilon_2} \cdot \delta_{k_1,k_2}.
\eeq
Since one easily finds that the determinant in Eq.~(\ref{eq:EffActionRewrittenForSmallY})
is invariant under the permutation $p\leftrightarrow\wp$, one can extend the 
product in that equation, which is performed only over one half of the whole
momentum space, again to the full momentum space $\ImpSpace$ by factorizing 
out the identity $\delta_{k_1,k_2}$. One then obtains for the effective 
action
\bea
S_{eff}[\Phi]
&=& S_\Phi[\Phi] - N_f\cdot \log\left[\prod\limits_{p\in\ImpSpace} {\det \left(\DnoK(p)\otimes\DnoK(\wp)
-\frac{\tilde y_N}{2\rho}\breve{\cal M}(p)\right)} \right],
\label{eq:EffActionRewrittenForSmallYCompactified}
\eea
with the definitions
\beq
\D(p) = \delta_{k_1,k_2} \cdot \DnoK(p), \quad
{\cal M}(p) = \delta_{k_1,k_2} \cdot\breve{\cal M}(p), \quad \mbox{and} \quad
{\cal M}^{a,b}(p) = \delta_{k_1,k_2} \cdot\breve{\cal M}^{a,b}(p),
\eeq
where $a,b\in \{1,2\}$. Selecting a special order for the indices $\zeta\epsilon$
according to $\{++,+-,-+,--\}$ the latter four $4\times 4$ matrices are 
explicitly given by
\bea
\breve{\cal M}^{1,1}(p)
&=& m\cdot
\left(
\begin{array}{*{4}{c}}
\hat\Phi^0\omega^+(p)&i\hat\Phi^1\omega^-(p)&0&(i\hat\Phi^3-\hat\Phi^2)\omega^-(p)\\
i\hat\Phi^1\omega^+(p)&\hat\Phi^0\omega^-(p)&(i\hat\Phi^3-\hat\Phi^2)\omega^+(p)&0\\
0&(i\hat\Phi^3+\hat\Phi^2)\omega^-(p)&\hat\Phi^0\omega^+(p)&-i\hat\Phi^1\omega^-(p)\\
(i\hat\Phi^3+\hat\Phi^2)\omega^+(p)&0&-i\hat\Phi^1\omega^+(p)&\hat\Phi^0\omega^-(p)\\
\end{array}
\right)\\
\breve{\cal M}^{1,2}(p) 
&= &s \cdot
\left(
\begin{array}{*{4}{c}}
i\hat\Phi^1\omega^+(\wp)&\hat\Phi^0\omega^-(\wp)&(i\hat\Phi^3-\hat\Phi^2)\omega^+(\wp)&0\\
\hat\Phi^0\omega^+(\wp)&i\hat\Phi^1\omega^-(\wp)&0&(i\hat\Phi^3-\hat\Phi^2)\omega^-(\wp)\\
(i\hat\Phi^3+\hat\Phi^2)\omega^+(\wp)&0&-i\hat\Phi^1\omega^+(\wp)&\hat\Phi^0\omega^-(\wp)\\
0&(i\hat\Phi^3+\hat\Phi^2)\omega^-(\wp)&\hat\Phi^0\omega^+(\wp)&-i\hat\Phi^1\omega^-(\wp)\\
\end{array}
\right)
\eea
where the abbreviation $\omega^\epsilon(p) = \nu^\epsilon(p)-2\rho$ was used.
The remaining matrices $\breve{\cal M}^{2,2}(p)$ and $\breve{\cal M}^{2,1}(p)$
are obtained from $\breve{\cal M}^{1,1}(p)$, $\breve{\cal M}^{1,2}(p)$ by 
interchanging $p$ and $\wp$. Using some algebraic manipulation package, the  
determinant of the $8\times 8$ matrix in Eq.~(\ref{eq:EffActionRewrittenForSmallYCompactified}) 
can be computed leading to the final expression for the effective action
\bea
S_{eff}[\Phi]
= S_\Phi[\Phi] - N_f\cdot\sum\limits_{p\in\ImpSpace}
\log&\Bigg[&\left(\left|\nu^+(p)\right|\cdot \left|\nu^+(\wp)\right| 
+\frac{\tilde y_N^2}{4\rho^2} \left(m^2 - s^2\right) \cdot 
\left|\nu^+(p)-2\rho\right| \cdot \left|\nu^+(\wp)-2\rho \right|\right)^2\nonumber \\
&+&m^2\frac{\tilde y_N^2}{4\rho^2}  
\Big(\left|\nu^+(p)-2\rho\right|\cdot \left|\nu^+(\wp) \right| 
- \left|\nu^+(\wp)-2\rho\right| \cdot \left|\nu^+(p) \right|\Big)^2 
\Bigg]^2. 
\label{eq:EffActionRewrittenForSmallYFinal}
\eea
With the ansatz in Eq.~(\ref{eq:staggeredAnsatz}) the Higgs field action $S_{\Phi}$
can also be written in terms of the quantities $m$ and $s$. One easily finds
\bea
S_{\Phi} 
&=& N_f \cdot L^4\cdot \Bigg\{-8\tilde\kappa_N \Big(m^2-s^2\Big) +  m^2+s^2 
+\tilde\lambda_N \Big( m^4 +s^4 + 6m^2s^2 -2\left(m^2+s^2 \right) \Big)\Bigg\}.
\eea
Two remarks are in order here for the orientation of the reader.

{\it (I)}
The resulting phase structure in the large $N_f$-limit can now be obtained by 
minimizing the effective action with respect to $m$ and $s$. In principle one 
could derive the corresponding phase diagrams for all values of the quartic coupling
constant $\tilde \lambda_N\ge 0$. However, as one can easily find from Eq.~(\ref{eq:DetOfScaleFacC})
the case $\tilde\lambda_N>0.5$ corresponds to the strong self-coupling regime $\lambda\gg 1$
of the physically underlying $\varphi^4$-theory given in Eq.~(\ref{eq:UsualPhiAction2})
for large values of $N_f$. In that regime it is no longer reasonable to evaluate
the effective action due to
the strong self-interaction of the Higgs-field in that case. We therefore restrict
the allowed range for the quartic coupling to $0\le\tilde\lambda_N < 0.5$, which
corresponds to the weak self-coupling regime of the physical model in Eq.~(\ref{eq:UsualPhiAction2}).

{\it (II)}
The sum over all allowed momenta $\ImpSpace$ in Eq.~(\ref{eq:EffActionRewrittenForSmallYFinal})
becomes a four-dimensional momentum integral over $\ImpSpace$ for $L=\infty$ according to
\beq
\frac{1}{L^4} \sum\limits_{p\in\ImpSpace}\; ... \quad \rightarrow \quad \int\limits_{p\in\ImpSpace} \frac{d^4p}{(2\pi)^4} \;...
\eeq
which was actually used in the numerical evaluation of the effective action.

\bc
\setlength{\unitlength}{0.01mm}
\begin{figure}[htb]
\begin{tabular}{cc}
$\tilde\lambda_N=0.1$ & $\tilde\lambda_N=0.3$ \\
\begin{picture}(6600,5500)
\put(600,500){\includegraphics[width=5cm]{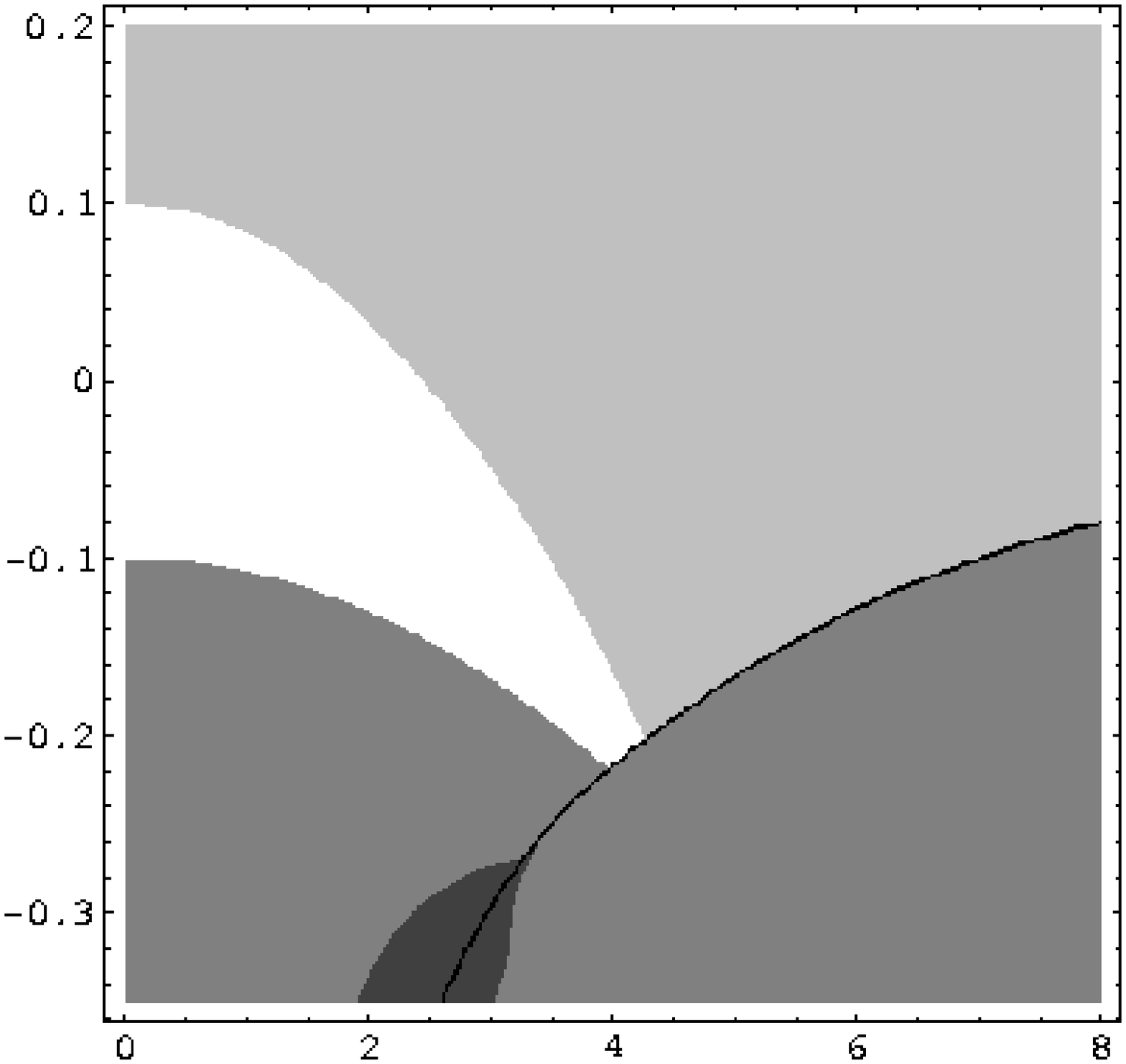}}
\put(25,3100){$\tilde \kappa_N$}
\put(3200,300){$\tilde y_N$}
\put(1300,3350){\textbf {SYM}}
\put(4500,3350){\textbf {FM}}
\put(1300,1500){\textbf {AFM}}
\put(4500,1500){\textbf {AFM}}
\put(2900,1000){$\leftarrow$\textbf {FI}}
\end{picture}
&
\begin{picture}(6600,5500)
\put(600,500){\includegraphics[width=5cm]{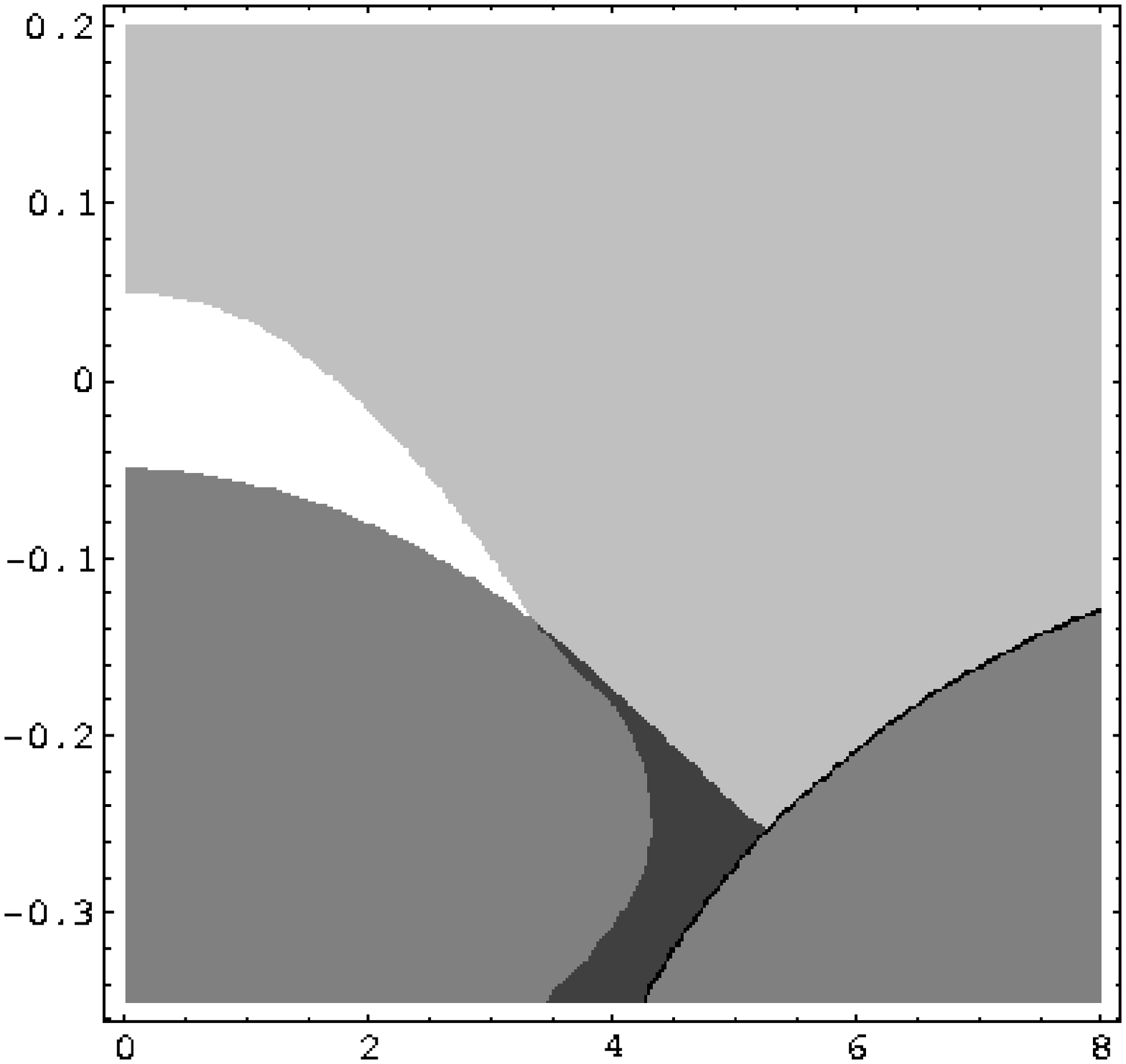}}
\put(25,3100){$\tilde \kappa_N$}
\put(3200,300){$\tilde y_N$}
\put(1300,3350){\textbf {SYM}}
\put(4500,3350){\textbf {FM}}
\put(1300,1500){\textbf {AFM}}
\put(4500,1500){\textbf {AFM}}
\put(2600,1000){\textbf {FI}$\rightarrow$}
\end{picture}
\\
\end{tabular}
\caption{Phase diagrams with respect to the Yukawa coupling constant $\tilde y_N$ and the hopping 
parameter $\tilde \kappa_N$ for the constant quartic couplings $\tilde \lambda_N=0.1$ (left)
and $\tilde \lambda_N=0.3$ (right). The black line indicates the first order phase transitions.
Both phase diagrams were determined for $L=\infty$.
An explanation of the occurring phases is given in the text.}
\label{fig:PhaseDiagrams1}
\end{figure}
\ec

We now present the phase diagrams for $\tilde\lambda_N=0.1$ and $\tilde\lambda_N=0.3$
in Fig.~\ref{fig:PhaseDiagrams1}. These phase diagrams were calculated for an infinite lattice,
\ie for $L=\infty$. Here we distinguish between the following four phases:\\
\begin{tabular}{crl}
&{\it (I)} &  The symmetric phase (SYM): $m=0,\, s=0$\\
&{\it (II)} & The ferromagnetic phase (FM): $m\neq 0,\, s=0$   \\
&{\it (III)} & The anti-ferromagnetic phase (AFM): $m=0,\, s\neq 0$  \\
&{\it (IV)} & The ferrimagnetic phase (FI): $m\neq 0,\, s\neq 0$   \\
\end{tabular}\\
In both cases, \ie $\tilde\lambda_N=0.1$ and $\tilde\lambda_N=0.3$,
one finds a symmetric phase approximately centered around $\tilde\kappa_N=0$ 
at sufficiently small values of the Yukawa coupling constant $\tilde y_N$,
as one would have expected, since the model becomes the pure $\phi^4$-theory
in the limit $\tilde y_N\rightarrow 0$. From the same consideration one would 
also expect the accompanying phase transitions to be of second order. This
is indeed the case as can clearly be seen in Fig.~\ref{fig:ExpectedMSPlots1}
showing the expectation values of the amplitudes $m$ and $s$ for different 
values of $\tilde y_N$ as obtained in the minimization process. With increasing 
$\tilde y_N$ the symmetric phase bends downwards to negative values of the 
hopping parameter $\tilde \kappa_N$, unless it either hits a first order phase 
transition to an anti-ferromagnetic phase (black line in Fig.~\ref{fig:PhaseDiagrams1}), 
the order of which can be determined from Fig.~\ref{fig:ExpectedMSPlots2} (this is 
the case for $\tilde\lambda_N=0.1$), or it eventually goes over into two FM-FI 
and FI-AFM second order phase transitions, which is the case for $\tilde\lambda_N=0.3$. 

Here we present only the expectation values 
of $m$ and $s$ for $\tilde \lambda_N=0.3$ and not for $\tilde \lambda_N=0.1$, since the 
latter plots would not provide qualitatively new information to the reader.

Interestingly, the ferrimagnetic phase (FI) exists in both presented scenarios, \ie
for $\tilde\lambda_N=0.1$ and $\tilde\lambda_N=0.3$, even 
deeply inside the anti-ferromagnetic phase region in the neighbourhood of the first
order phase transition boundary. However, due to the small expectation value of the
amplitude $m$ of the constant mode (see Fig.~\ref{fig:ExpectedMSPlots1}) it is
questionable whether this ferrimagnetic phase will be observable in corresponding 
numerical simulations.

\bc
\setlength{\unitlength}{0.01mm}
\begin{figure}[htb]
\begin{picture}(12000,13144)
\put(0,1000){\includegraphics[angle=0,width=0.27\textwidth]{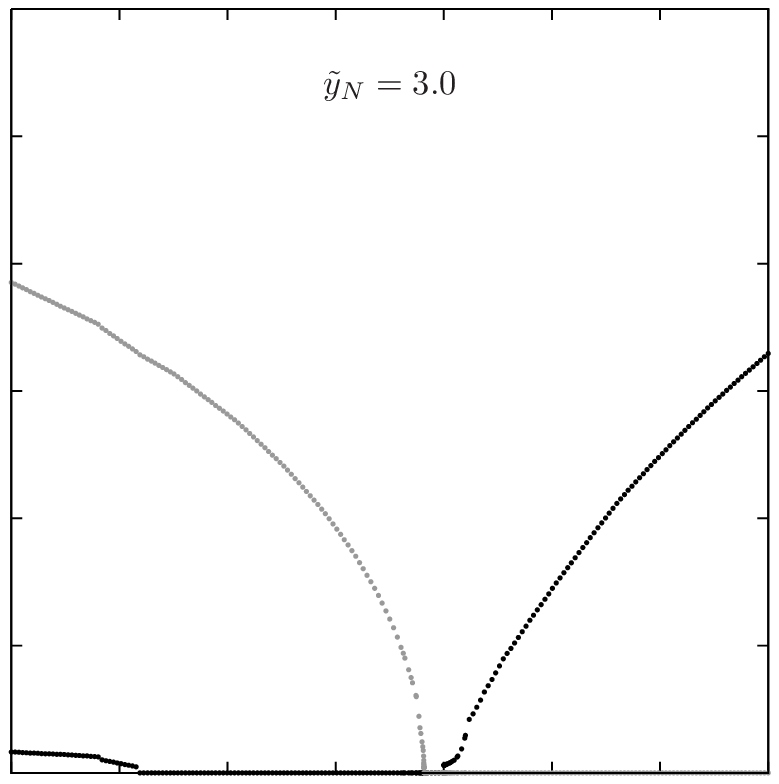}}
\put(4000,1000){\includegraphics[angle=0,width=0.27\textwidth]{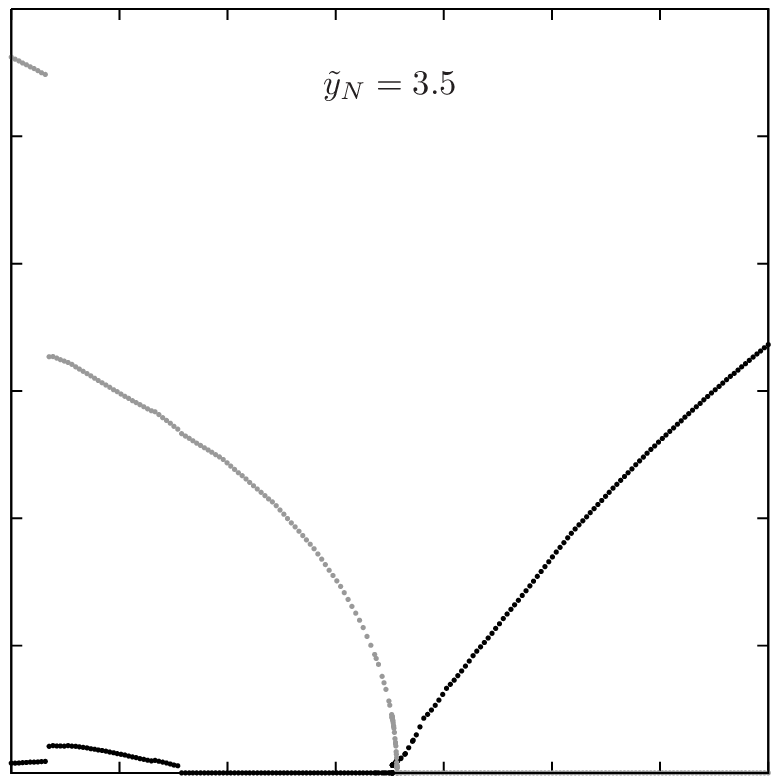}}
\put(8000,1000){\includegraphics[angle=0,width=0.27\textwidth]{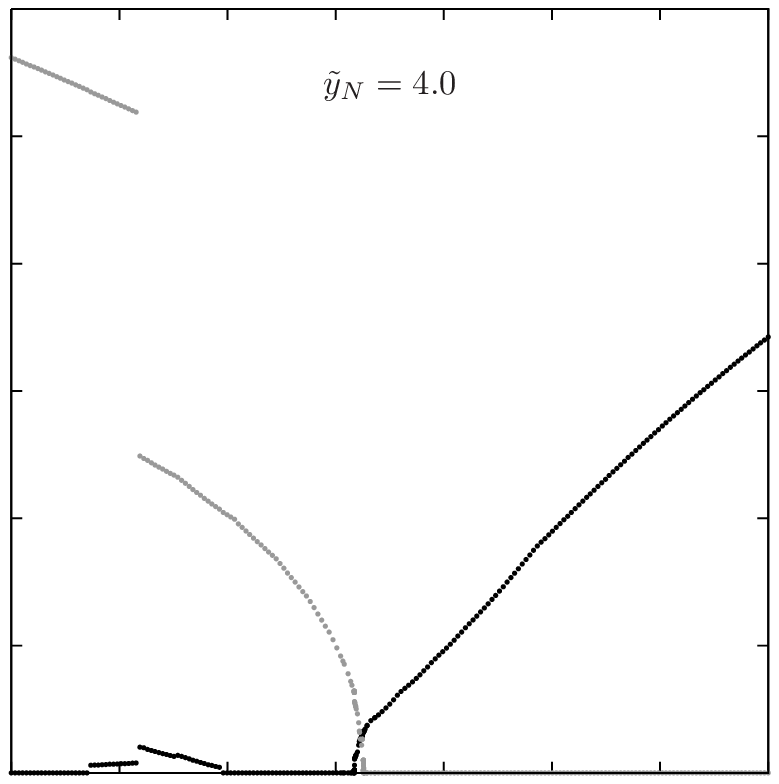}}
\put(0,5048){\includegraphics[angle=0,width=0.27\textwidth]{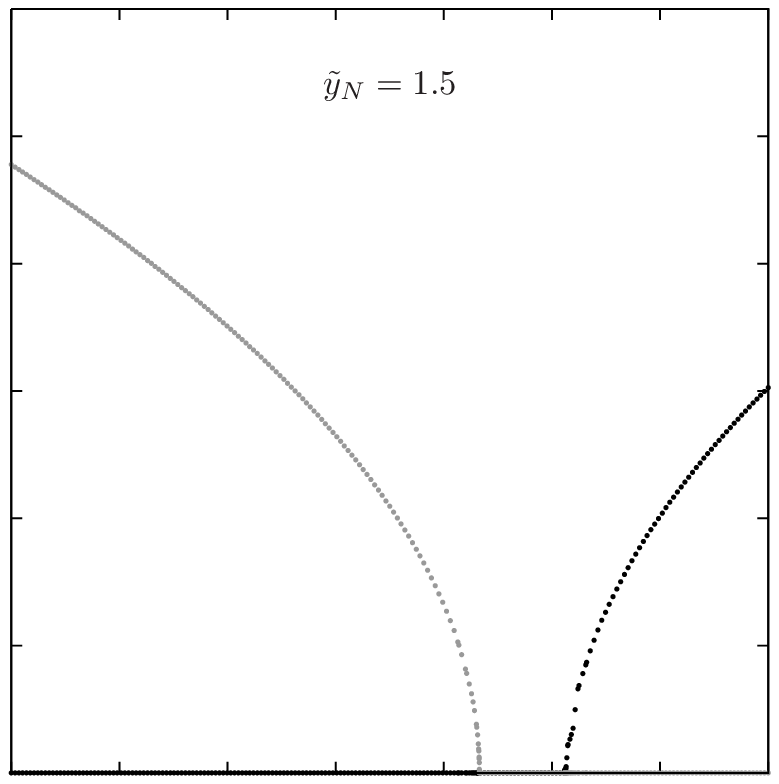}}
\put(4000,5048){\includegraphics[angle=0,width=0.27\textwidth]{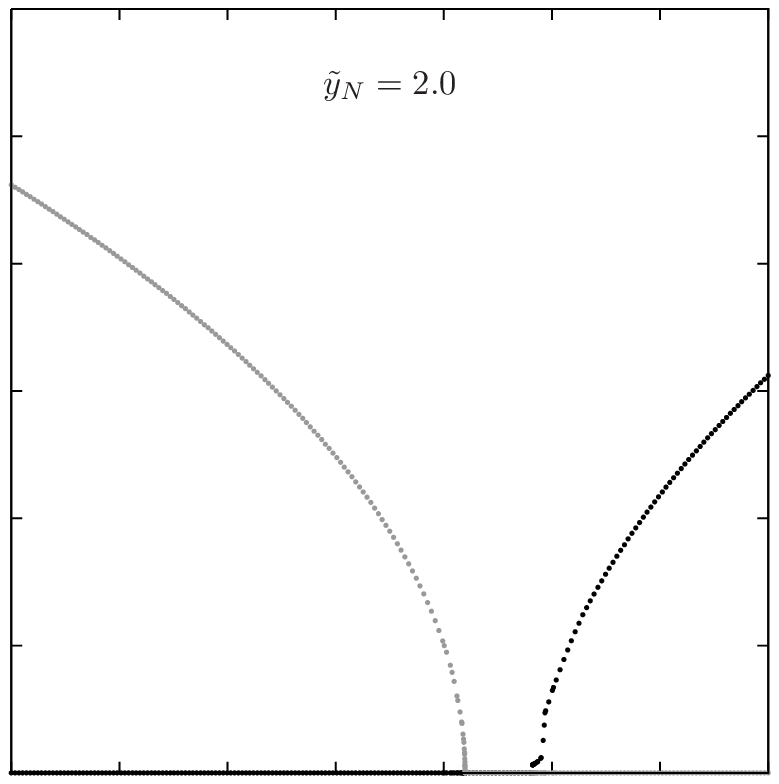}}
\put(8000,5048){\includegraphics[angle=0,width=0.27\textwidth]{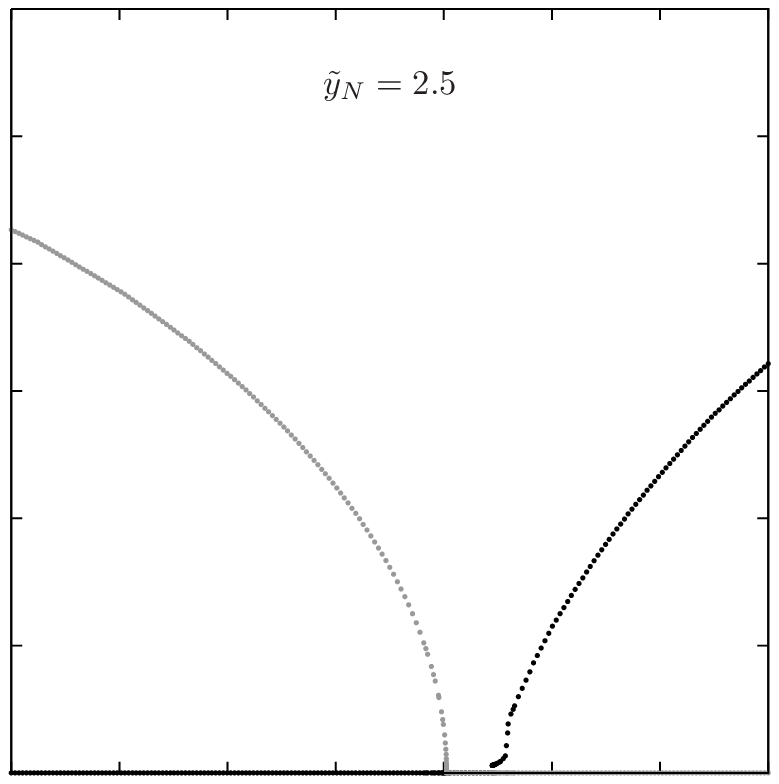}}
\put(0,9096){\includegraphics[angle=0,width=0.27\textwidth]{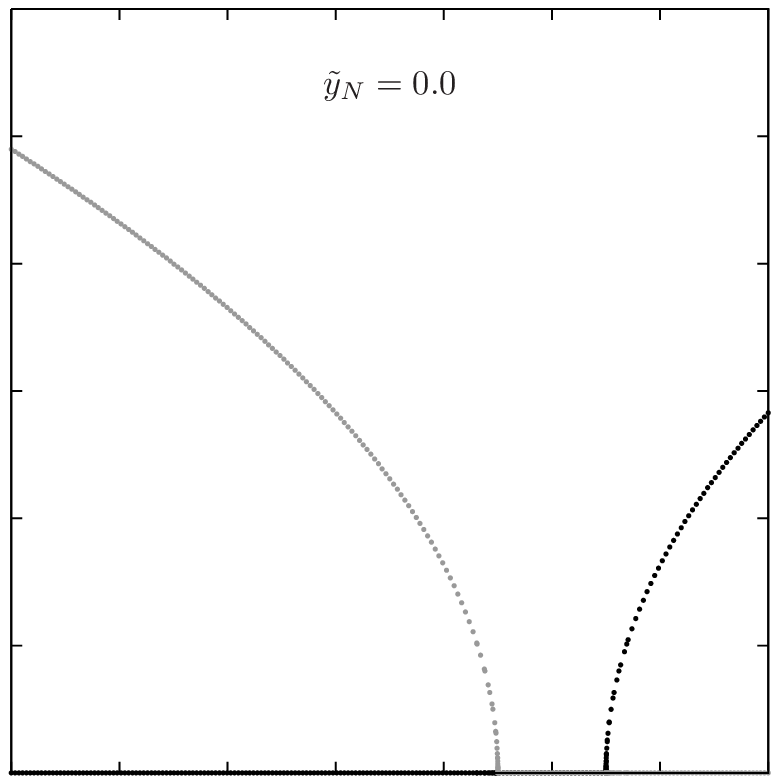}}
\put(4000,9096){\includegraphics[angle=0,width=0.27\textwidth]{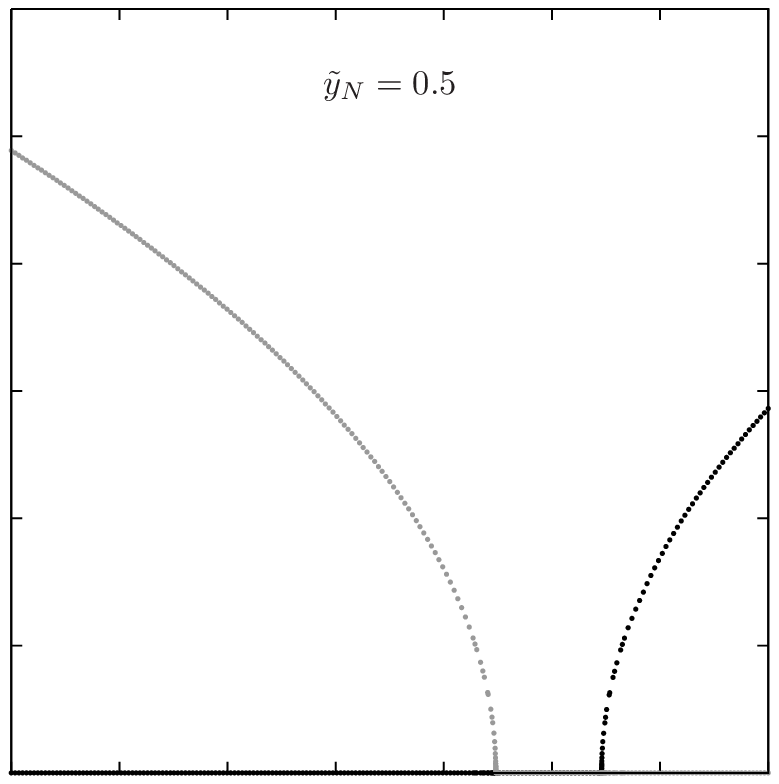}}
\put(8000,9096){\includegraphics[angle=0,width=0.27\textwidth]{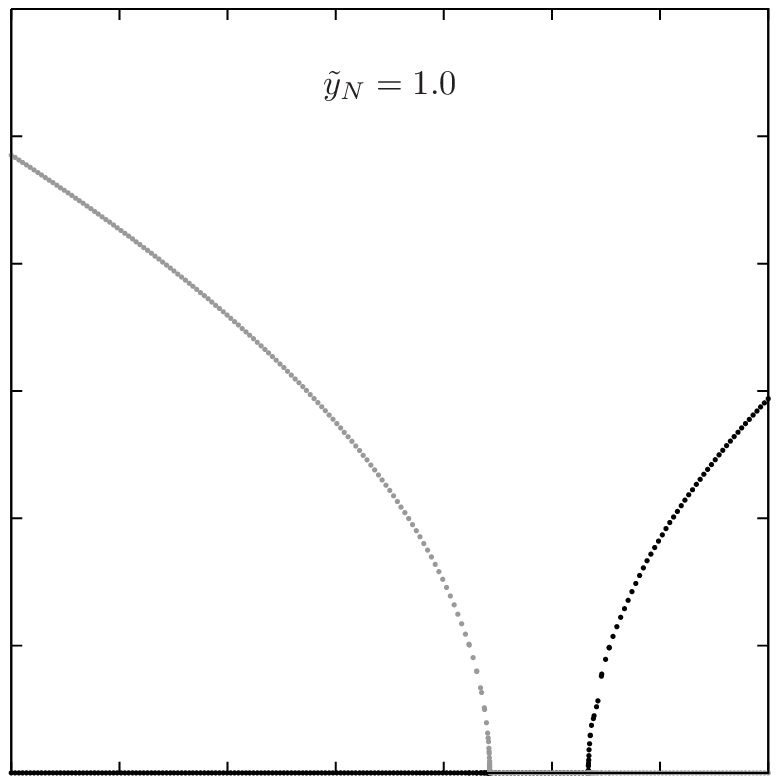}}

\put(-200,700){\tiny{-0.5}}
\put(+381,700){\tiny{-0.4}}
\put(+942,700){\tiny{-0.3}}
\put(+1513,700){\tiny{-0.2}}
\put(+2084,700){\tiny{-0.1}}
\put(+2735,700){\tiny{0.0}}
\put(+3316,700){\tiny{0.1}}

\put(3800,700){\tiny{-0.5}}
\put(4381,700){\tiny{-0.4}}
\put(4942,700){\tiny{-0.3}}
\put(5513,700){\tiny{-0.2}}
\put(6084,700){\tiny{-0.1}}
\put(6735,700){\tiny{0.0}}
\put(7316,700){\tiny{0.1}}

\put(7800,700){\tiny{-0.5}}
\put(8381,700){\tiny{-0.4}}
\put(8942,700){\tiny{-0.3}}
\put(9513,700){\tiny{-0.2}}
\put(10084,700){\tiny{-0.1}}
\put(10735,700){\tiny{0.0}}
\put(11316,700){\tiny{0.1}}
\put(11870,700){\tiny{0.2}}

\put(5800,100){$\tilde \kappa_N$}

\put(-430,960){\tiny{0.0}}
\put(-430,1640){\tiny{0.5}}
\put(-430,2320){\tiny{1.0}}
\put(-430,3000){\tiny{1.5}}
\put(-430,3670){\tiny{2.0}}
\put(-430,4340){\tiny{2.5}}

\put(-430,5008){\tiny{0.0}}
\put(-430,5688){\tiny{0.5}}
\put(-430,6368){\tiny{1.0}}
\put(-430,7048){\tiny{1.5}}
\put(-430,7718){\tiny{2.0}}
\put(-430,8388){\tiny{2.5}}

\put(-430,9056){\tiny{0.0}}
\put(-430,9736){\tiny{0.5}}
\put(-430,10416){\tiny{1.0}}
\put(-430,11096){\tiny{1.5}}
\put(-430,11766){\tiny{2.0}}
\put(-430,12436){\tiny{2.5}}
\put(-430,13104){\tiny{3.0}}
\end{picture}
\caption{Expectation values for the amplitudes of the constant ($m$: black curve) and staggered ($s$: gray curve) modes 
for several selected values of the Yukawa coupling constant $\tilde y_N$ and a constant quartic coupling $\tilde \lambda_N=0.3$.
The results were obtained for $L=\infty$.}
\label{fig:ExpectedMSPlots1}
\end{figure}
\ec

\bc
\setlength{\unitlength}{0.01mm}
\begin{figure}[htb]
\begin{picture}(12000,13144)
\put(0,1000){\includegraphics[angle=0,width=0.27\textwidth]{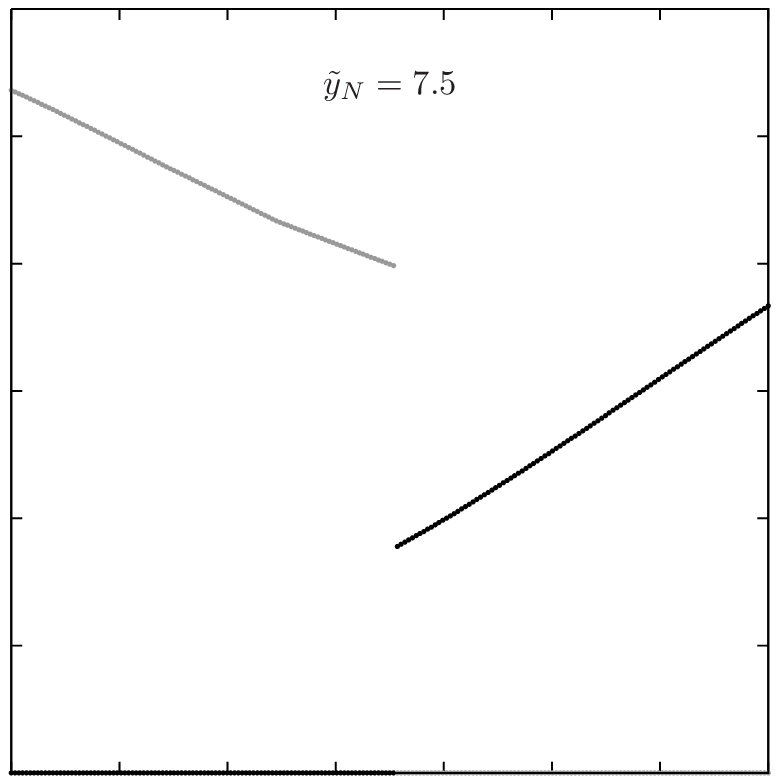}}
\put(4000,1000){\includegraphics[angle=0,width=0.27\textwidth]{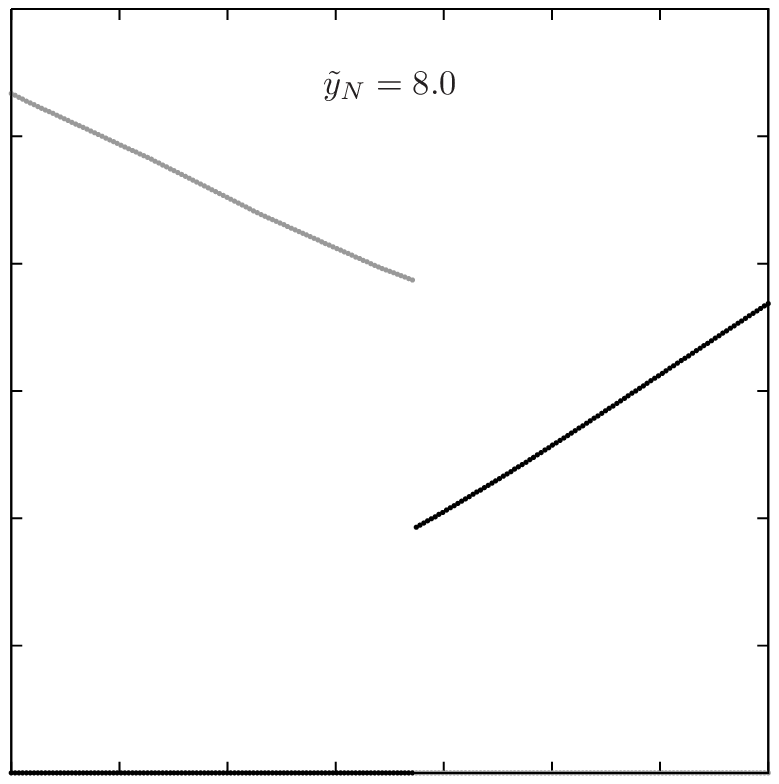}}
\put(8000,1000){\includegraphics[angle=0,width=0.27\textwidth]{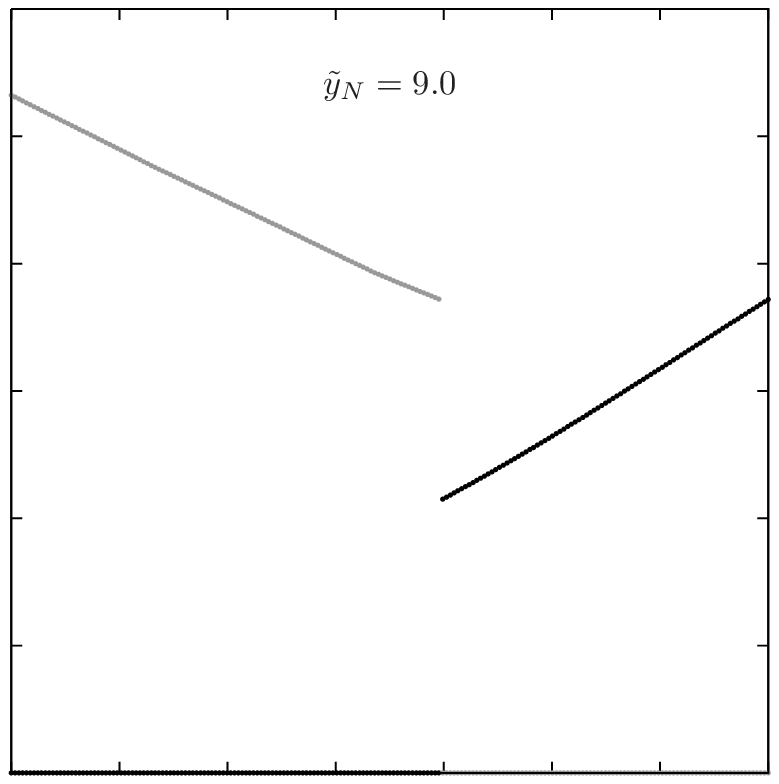}}
\put(0,5048){\includegraphics[angle=0,width=0.27\textwidth]{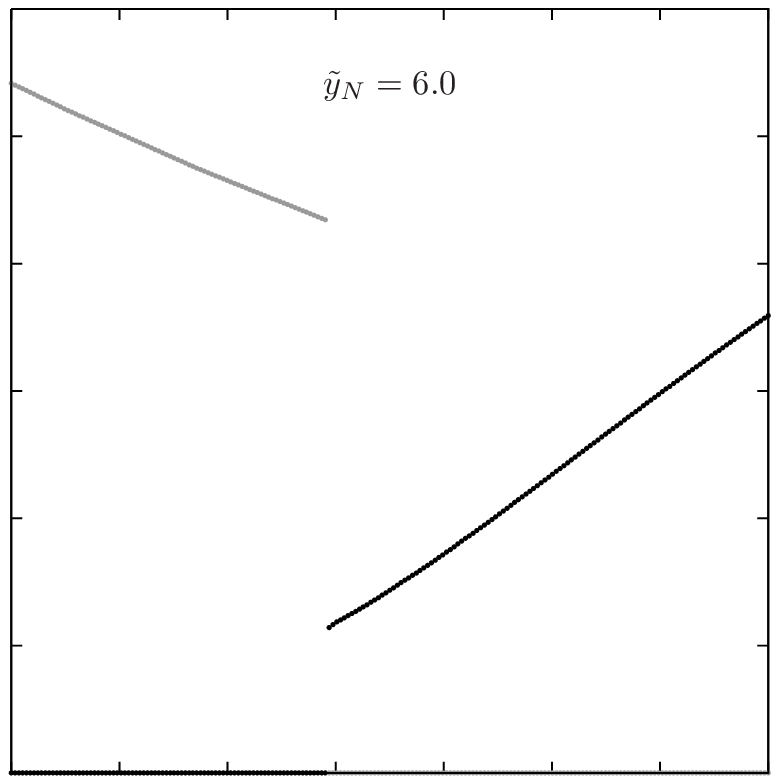}}
\put(4000,5048){\includegraphics[angle=0,width=0.27\textwidth]{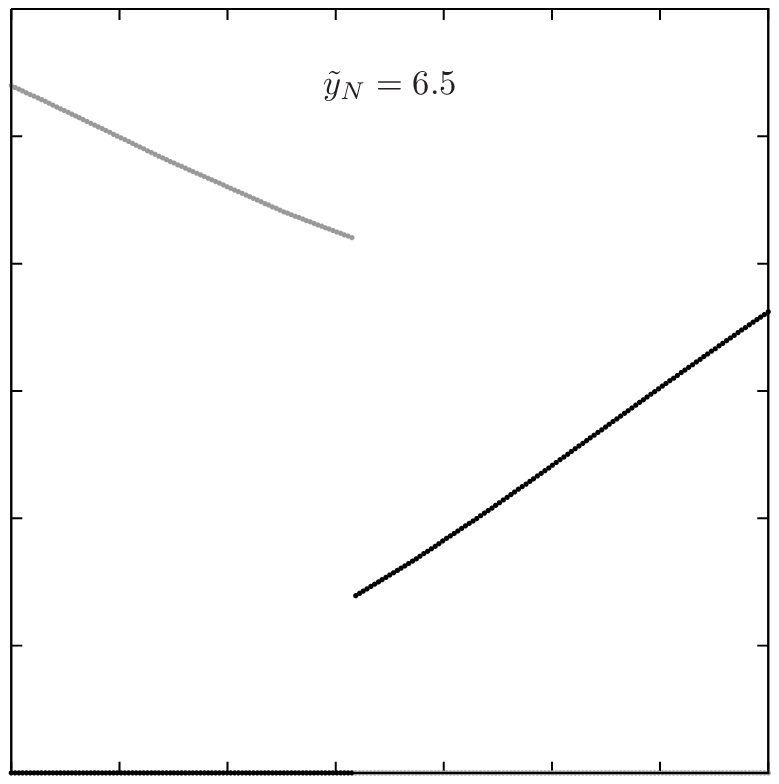}}
\put(8000,5048){\includegraphics[angle=0,width=0.27\textwidth]{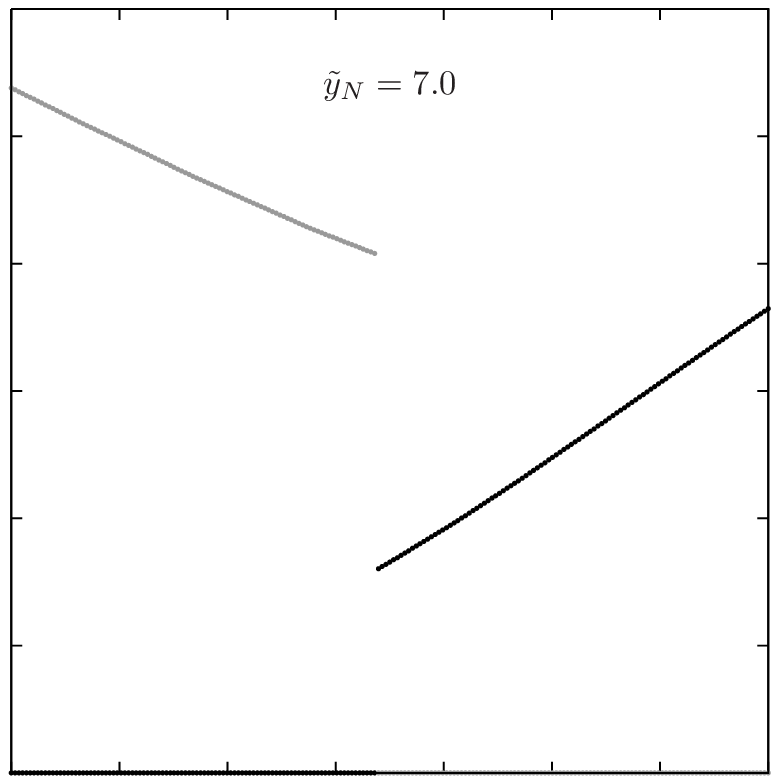}}
\put(0,9096){\includegraphics[angle=0,width=0.27\textwidth]{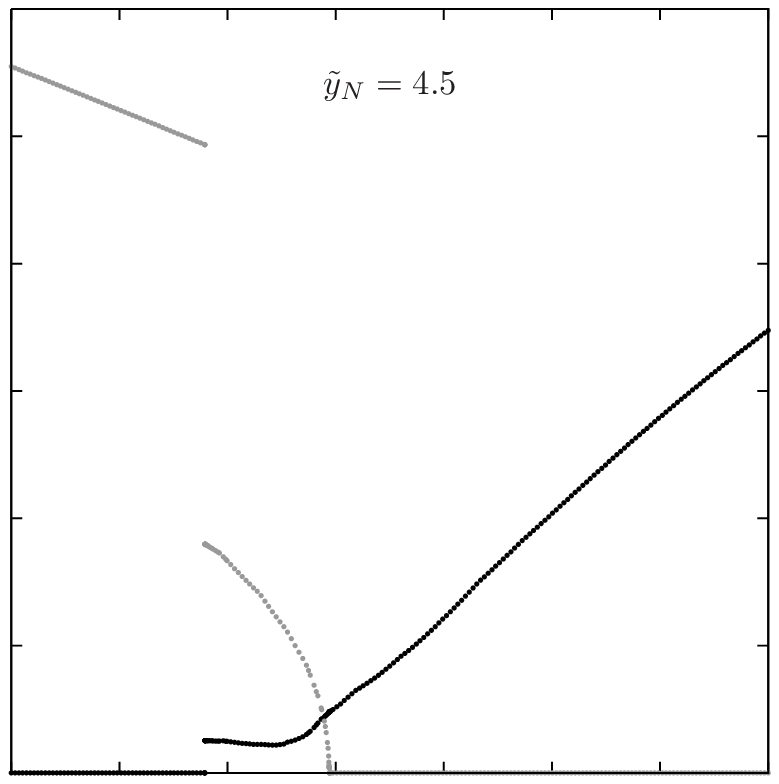}}
\put(4000,9096){\includegraphics[angle=0,width=0.27\textwidth]{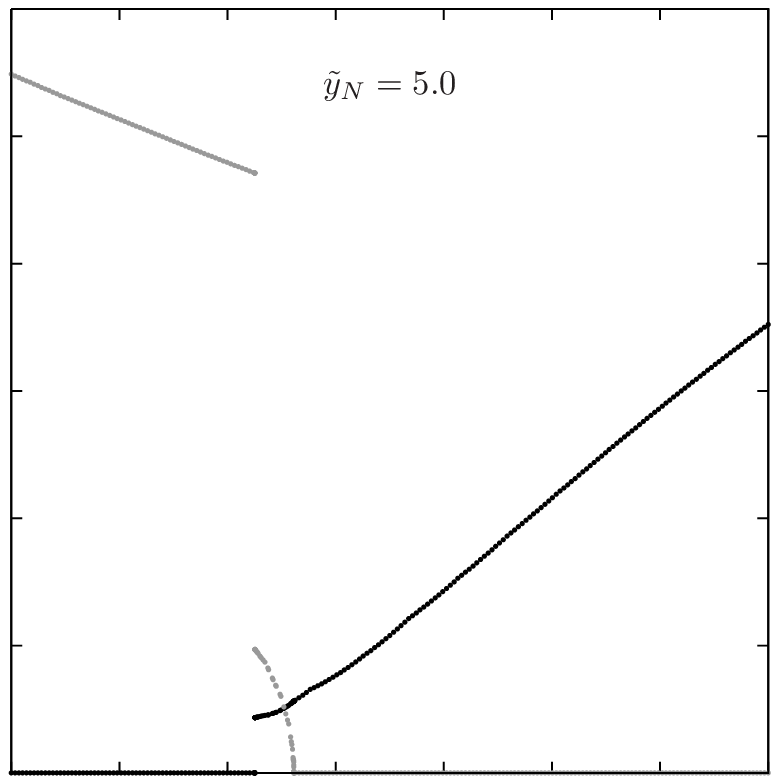}}
\put(8000,9096){\includegraphics[angle=0,width=0.27\textwidth]{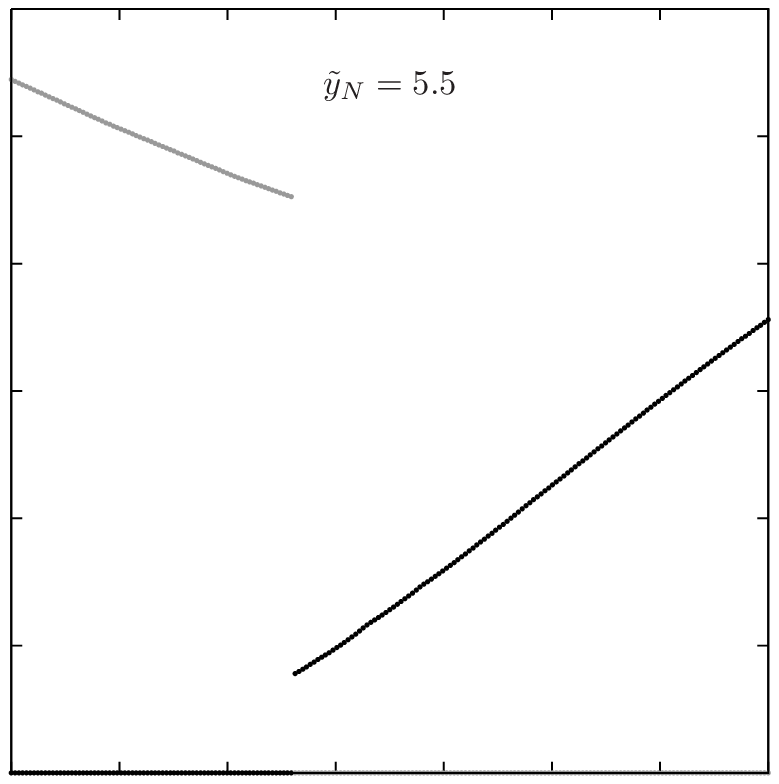}}

\put(-200,700){\tiny{-0.5}}
\put(+381,700){\tiny{-0.4}}
\put(+942,700){\tiny{-0.3}}
\put(+1513,700){\tiny{-0.2}}
\put(+2084,700){\tiny{-0.1}}
\put(+2735,700){\tiny{0.0}}
\put(+3316,700){\tiny{0.1}}

\put(3800,700){\tiny{-0.5}}
\put(4381,700){\tiny{-0.4}}
\put(4942,700){\tiny{-0.3}}
\put(5513,700){\tiny{-0.2}}
\put(6084,700){\tiny{-0.1}}
\put(6735,700){\tiny{0.0}}
\put(7316,700){\tiny{0.1}}

\put(7800,700){\tiny{-0.5}}
\put(8381,700){\tiny{-0.4}}
\put(8942,700){\tiny{-0.3}}
\put(9513,700){\tiny{-0.2}}
\put(10084,700){\tiny{-0.1}}
\put(10735,700){\tiny{0.0}}
\put(11316,700){\tiny{0.1}}
\put(11870,700){\tiny{0.2}}

\put(5800,100){$\tilde \kappa_N$}

\put(-430,960){\tiny{0.0}}
\put(-430,1640){\tiny{0.5}}
\put(-430,2320){\tiny{1.0}}
\put(-430,3000){\tiny{1.5}}
\put(-430,3670){\tiny{2.0}}
\put(-430,4340){\tiny{2.5}}

\put(-430,5008){\tiny{0.0}}
\put(-430,5688){\tiny{0.5}}
\put(-430,6368){\tiny{1.0}}
\put(-430,7048){\tiny{1.5}}
\put(-430,7718){\tiny{2.0}}
\put(-430,8388){\tiny{2.5}}

\put(-430,9056){\tiny{0.0}}
\put(-430,9736){\tiny{0.5}}
\put(-430,10416){\tiny{1.0}}
\put(-430,11096){\tiny{1.5}}
\put(-430,11766){\tiny{2.0}}
\put(-430,12436){\tiny{2.5}}
\put(-430,13104){\tiny{3.0}}
\end{picture}
\caption{Expectation values for the amplitudes of the constant ($m$: black curve) and staggered ($s$: gray curve) modes 
for several selected values of the Yukawa coupling constant $\tilde y_N$ and a constant quartic coupling $\tilde \lambda_N=0.3$.
The results were obtained for $L=\infty$.}
\label{fig:ExpectedMSPlots2}
\end{figure}
\ec

\section{Large $N_f$-limit for large Yukawa coupling parameters}
\label{sec:LargeYukawaCouplings}
In this section we will examine the phase diagram of the considered 
Higgs-Yukawa model in the regime of large values of the Yukawa coupling 
constant $y_N$ and for arbitrary values of the quartic coupling constant
$\lambda_N\ge 0$. This will be done in three steps. Firstly, the effective
action is expanded in powers of the inverse coupling constant $1/y_N$. 
Taking only the first non-vanishing contribution of this power series
into account and performing the large $N_f$-limit in such a way,
that the amplitude of the Higgs field is fixed, the model then effectively
becomes an $O(4)$-symmetric, non-linear sigma-model up to some finite
volume terms. Finally, the phase diagram of the latter sigma-model is 
determined by an additional large $N$-limit, where $N$ denotes here
the number of Higgs field components. 

For an evaluation of the effective action it is crucial to pay 
special attention to the fermion doubler modes
\beq
\ImpBasis_\pi = \left\{\Psi^{p,\zeta\epsilon k} : p_\mu\in\{0,\pi\},\, p\neq 0,\, \zeta,\epsilon=\pm 1,\, k\in\{1,2\} \right\}
\label{eq:DefPiModes}
\eeq
which we will refer to as $\pi$-modes in the following. Given these $120$ modes one can define the 
corresponding projection operator
\beq
P_\pi = \sum\limits_{\Psi \in \ImpBasis_\pi} \Psi \Psi^\dagger
\label{eq:DefOfProjector}
\eeq
projecting to the sub-space $V_\pi=\mbox{span}(\ImpBasis_\pi)$ spanned by $\ImpBasis_\pi$. 
Using this notation one can easily establish the very helpful relation
\bea
\label{eq:DetRelationWithProjector}
\det\left(E\left(\ID-P_\pi\right) + P_\pi F P_\pi \right)
&=&
\det\left(\left(\ID-P_\pi\right)E\left(\ID-P_\pi\right) + P_\pi F P_\pi \right) \nonumber \\
&=&
\det\left(\left(\ID-P_\pi\right)E + P_\pi F P_\pi \right)
=
\mbox{det}'\left( E \right) \cdot \mbox{det}^*\left( F \right)
\eea
where $E$ and $F$ are arbitrary operators defined on the same space $V$ as $\D$ and $B$. Here the expression
$\mbox{det}^*\left( F \right)$ denotes the determinant of $F$ with respect to the sub-space $V_\pi$ and
$\mbox{det}'\left( E \right)$ is the determinant of $E$ with respect to the complementary space 
$V/V_\pi\equiv \mbox{span}(\ImpBasis / \ImpBasis_\pi)$, where $\ImpBasis$ denotes the full
set of all modes. 
Using Eq.~(\ref{eq:DetRelationWithProjector}) several times one can rewrite the effective action according to
\bea
\label{eq:ReductionOfEffectiveAction}
e^{-\frac{S_{eff}[\Phi]-S_{\Phi}}{N_f}} 
&=&
\det\left( y_NB\left(\D-2\rho\right)-2\rho\D\right) \\
&=&
\left(-4\rho^2\right)^{120} \cdot \mbox{det}'\left( y_NB'\left(\Dprime-2\rho\ID'\right)-2\rho\Dprime  \right) \nonumber \\
&=&
\left(-4\rho^2\right)^{120} \cdot \mbox{det}'\left( y_N\right) \cdot \mbox{det}'\left( \Dprime-2\rho\ID'\right) \cdot
\mbox{det}'\left( B'-\frac{2\rho}{y_N}\Dprime\left(\Dprime-2\rho\ID'\right)^{-1}  \right) \nonumber \\
&=&
\mbox{Const} \cdot
\det\left(  B - \left(B-\ID\right)P_\pi -\frac{2\rho}{y_N} \DDmtwoRhoInv   \right) \nonumber \\
&=&
\mbox{Const} \cdot \det\left(B\right) \cdot \det\left(\ID-\left(\ID-B^{-1}\right)P_\pi\right)
\cdot \det\left(\ID - \frac{2\rho}{y_N}B^{-1}\DDmtwoRhoInv \left[ \ID - \left(\ID-B^{-1}\right)P_\pi \right]^{-1} \right)
\nonumber
\eea
where $\Dprime$, $B'$, and $\ID'$ denote the restriction of the operators $\D$, $B$, and $\ID$ to
the sub-space $V/V_\pi$. This restriction is introduced, since it guarantees $\Dprime-2\rho\ID'$ to be 
invertible. The operator $\DDmtwoRhoInv$ is then defined by extending the domain of the inverse of 
$\Dprime-2\rho\ID'$ again to the full space $V$ by inserting the projector 
$\ID-P_\pi$ according to
\beq
\label{eq:DefOfOperatorA}
\DDmtwoRhoInv = \Dprime\cdot \left[\Dprime-2\rho\ID'  \right]^{-1} \cdot 
\left(\ID - P_\pi\right),
\eeq
which is well-defined and finite over the whole space $V$. The last 
determinant in Eq.~(\ref{eq:ReductionOfEffectiveAction}) can further be 
reduced by using the result
\bea
\left[ \ID - \left(\ID-B^{-1}\right)P_\pi \right]^{-1} &=&
\ID - P_\pi + P_\pi\left(1-P_\pi + P_\pi B^{-1} P_\pi \right)^{-1}P_\pi \nonumber \\
&-&\left(\ID-P_\pi \right) B^{-1} P_\pi \left(1-P_\pi + P_\pi B^{-1} P_\pi \right)^{-1}P_\pi
\eea
and by applying again relation~(\ref{eq:DetRelationWithProjector}) leading to the
compact notation for the effective action
\bea
\label{eq:EffectiveActionFullLargeYFirst}
S_{eff}[\Phi] \;=\; S_\Phi &-& N_f \cdot \log\det\left(B\right) 
- N_f \cdot \log\mbox{det}^*\left(B^{-1}\right) 
- N_f \cdot \log\mbox{det}^*\left(\ID + \frac{2\rho}{y_N} F[\Phi] \right)  \nonumber\\
&-& N_f \cdot \log\det\left(\ID - \frac{2\rho}{y_N} \DDmtwoRhoInv
\cdot B^{-1} \right) ,
\eea
with the abbreviation $F[\Phi]$ defined as the somewhat lengthy expression
\bea
\label{eq:DefOfFiniteTermMat}
F[\Phi] &=&  \left[\ID-\frac{2\rho}{y_N} B^{-1}\DDmtwoRhoInv\right]^{-1}
B^{-1} \DDmtwoRhoInv B^{-1} P_\pi \left[\ID-P_\pi +P_\pi B^{-1}  P_\pi   \right]^{-1}.
\eea
However, the latter determinants $\mbox{det}^*$ only give rise to some finite volume effects, 
since these determinants are only performed over the 120-dimensional sub-space $V_\pi$. Their 
contributions to the effective action do therefore {\it not} scale proportional to $L^4$ as the lattice 
size increases in contrast to all other appearing terms.
We will come back to discussing these finite volume effects later. 
Here, we will first continue with the evaluation of the last term in 
Eq.~(\ref{eq:EffectiveActionFullLargeYFirst}) by rewriting the corresponding trace 
as a power series in the inverse coupling constant $1/y_N$ according to
\bea
\label{eq:DefPowerSeries}
\mbox{Tr}\log\left(\ID - \frac{2\rho}{y_N} \DDmtwoRhoInv \cdot B^{-1} \right)
&=&
- \mbox{Tr}\, \sum\limits_{r=1}^{\infty} \frac{2^r}{r} \left(\frac{\rho}{y_N}\right)^r \left[ \DDmtwoRhoInv B^{-1}\right]^r 
\eea
and by eventually cutting off this power series after the first
non-vanishing term, which is well-justified for sufficiently large $y_N$.
For our purpose of establishing the desired connection to a sigma-model
it is most convenient to evaluate these expressions in position space. 
Here the matrix $B^{-1}$ is block diagonal and explicitly given by
\beq
\label{eq:InverseOfB}
B^{-1} = B^\dagger \cdot \left(BB^\dagger  \right)^{-1}, \quad
B^{-1}_{m,n} = \delta_{m,n} \cdot \hat B(\Phi_n^*/|\Phi_n|^2),
\eeq
where the notation $(\Phi^*_n)^0=\Phi^0_n$, $(\Phi^*_n)^i= -\Phi^i_n$
was used and $\hat B$ was defined in Eq.~(\ref{eq:DefBHat}). 
In position space the matrix $\DDmtwoRhoInv B^{-1}$ can hence be written as
\bea
\label{eq:CalcOfABInv}
\left[\DDmtwoRhoInv B^{-1}\right]_{n_1,n_2} &=& \sum\limits_{p\in\ImpSpace} \sum\limits_{\zeta\epsilon k}  \frac{e^{ipn_1} u^{\zeta\epsilon k}(p) \alpha^\epsilon(p)
e^{-ipn_2}\left[u^{\zeta\epsilon k}(p)\right]^\dagger}{|\Psi^{p,\zeta\epsilon k}|^2} \hat B(\Phi^*_{n_2}/|\Phi_{n_2}|^2)  \\
&=& \frac{1}{L^4}\sum\limits_{p\in\ImpSpace} \sum\limits_{{\zeta\epsilon k}\atop{\zeta'\epsilon' k'}}  \alpha^\epsilon(p) e^{ip(n_1-n_2)}
{u^{\zeta\epsilon k}(p)}  \left(\hat B^{(p)}(\Phi^*_{n_2}/|\Phi_{n_2}|^2)\right)_{\zeta\epsilon k, \zeta'\epsilon' k'}  
{\left[u^{\zeta'\epsilon' k'}(p)\right]^\dagger} \nonumber
\eea
with $\hat B^{(p)}$ as defined in Eq.~(\ref{eq:MatBinSpinorRep}). The scalars $\alpha^\epsilon(p)$ 
denote the eigenvalues of the anti-hermitian operator $\DDmtwoRhoInv$ corresponding to its eigenvectors 
$\Psi^{p,\zeta\epsilon k}$ and are explicitly given by
\bea
\label{eq:defOfAlpha}
i\re\ni \alpha^\epsilon(p) &=& 
\Bigg\{
\begin{array}{*{3}{ccl}}
\frac{\nu^\epsilon(p)}{\nu^\epsilon(p) - 2\rho}  &:& p\in\ImpSpace,\,\nu^\epsilon(p)\neq 2\rho        \\ 
0  &:& p\in\ImpSpace,\,\nu^\epsilon(p) =  2\rho\\
\end{array}.
\eea
The result for the trace of the operator $\DDmtwoRhoInv B^{-1}$ is then directly found to be
\bea
\label{eq:TrGnPower1}
Tr\, \left[ \DDmtwoRhoInv B^{-1} \right] 
&=& \frac{1}{L^4} \sum\limits_{n}\sum\limits_{p\in\ImpSpace} \mbox{Tr}_{8\times 8} \left[
|\Phi_{n}|^{-2} \DDmtwoRhoInv(p)
\hat B^{(p)}(\Phi^*_{n})\right],
\eea
which can be generalized to the trace of the $r$-th power of $\DDmtwoRhoInv B^{-1}$ yielding
\bea
\label{eq:TrGn1}
Tr\, \left[ \DDmtwoRhoInv B^{-1} \right]^r 
&=& \sum\limits_{{n_1,...,n_r}\atop{p_1,...,p_r\in\ImpSpace}} 
Tr_{8\times 8}\, \left[  \prod\limits_{i=1}^r \frac{e^{ip_i(n_i-n_{i+1})}}{L^4} 
|\Phi_{n_{i+1}}|^{-2} \DDmtwoRhoInv(p_i)  \left(\hat B^{(p_i)}(\Phi^*_{n_{i+1}})\right) U(p_i,p_{i+1}) \right]
\eea
where $p_{n+1}$ is identified with $p_1$, and $x_{n+1}$ with $x_1$, and 
the expression $\DDmtwoRhoInv(p)$ stands for the diagonal matrix 
\beq
\DDmtwoRhoInv(p)_{\zeta_1\epsilon_1 k_1,\zeta_2\epsilon_2 k_2} = 
\delta_{\zeta_1,\zeta_2} \cdot \delta_{\epsilon_1,\epsilon_2} \cdot \delta_{k_1,k_2} \cdot
\alpha^{\epsilon_1}(p).
\eeq
At this point we refer the interested reader to Appendix~\ref{AppendixRthPowerTres}
for the details of this calculation in order to sustain the readability of this text.

However, it turns out that the evaluation of Eq.~(\ref{eq:TrGn1}) becomes 
much easier, if one inserts the identity $U(p_i,0)U(0,p_i)$ at some proper 
places. The remaining $8\times 8$ trace can then be simplified to
\bea
\label{eq:TrGn2}
Tr_{8\times 8}\, \left[  \prod\limits_{i=1}^r \DDmtwoRhoInv(p_i) \left(\hat B^{(p_i)}(\Phi^*_{n_{i+1}})\right) U(p_i,p_{i+1}) \right]
&=& 
Tr_{8\times 8}\, \left[\prod\limits_{i=1}^r \DDmtwoRhoInv^{(0)}(p_i) \left(\hat B^{(0)}(\Phi^*_{n_{i+1}})\right) \right],
\eea
where the representation $\hat B^{(0)}(\Phi^\dagger_n)$ of the Yukawa coupling matrix can 
directly be taken from Eq.~(\ref{eq:MatBinSpinorRep}) and $\DDmtwoRhoInv^{(0)}(p)$ is given by
\bea
\DDmtwoRhoInv^{(0)}(p) &=& U(0,p) \DDmtwoRhoInv(p) U(p,0) \nonumber \\
&=& \frac{\alpha^+(p)}{\sqrt{\tilde p^2}} \cdot 
\left(
\begin{array}{*{2}{c}}
\tilde p_0 & -\vec{\tilde p}\vec\Theta\\
\vec{\tilde p}\vec\Theta & -\tilde p_0\\
\end{array}
\right) 
\otimes
\left(
\begin{array}{*{2}{c}}
\tilde p_0 & -\vec{\tilde p}\vec\Theta\\
\vec{\tilde p}\vec\Theta & -\tilde p_0\\
\end{array}
\right) 
\eea
where the relation $\alpha^+(p) = - \alpha^-(p)$ has implicitly been used. 
Due to the insertion of the spinor basis transformation matrices $U(p_i,0)$
and $U(0,p_i)$ the sums over the momenta in Eq.~(\ref{eq:TrGn1}) factorize now according to
\bea
\label{eq:DefOfMatStrucT}
Tr\,\left[\DDmtwoRhoInv B^{-1}   \right]^r
&=& \sum\limits_{n_1,...,n_r} Tr_{8\times 8}\, \Bigg[  \prod\limits_{i=1}^r 
\underbrace{\left(\sum\limits_{p_i\in\ImpSpace} \frac{e^{ip_i(n_i-n_{i+1})}}{L^4}  \DDmtwoRhoInv^{(0)}(p_i) \right)  
|\Phi_{n_{i+1}}|^{-2}\left(\hat B^{(0)}(\Phi^*_{n_{i+1}})\right)}_{{\cal T}_{n_i,n_{i+1}}}  \Bigg] 
\eea
where each momentum sum is a four-dimensional Fourier transform of an anti-symmetric and purely imaginary 
summand, hence yielding real values. With the definition 
\beq
\label{eq:DefFourTransOfAlpha}
\re\ni \Gamma_\mu(\Delta n) = -\Gamma_\mu(-\Delta n) = 
\sum\limits_{p\in\ImpSpace} \frac{e^{ip\Delta n}}{L^4} \alpha^+(p) \cdot \frac{\tilde p_\mu}{\sqrt{\tilde p^2}} 
\eeq
the hermitian matrix ${\cal T}_{n,m}$ appearing in Eq.~(\ref{eq:DefOfMatStrucT}) can compactly be written
as
\beq
\label{eq:ExplicitTn}
{\cal T}_{n,m} = \frac{1}{|\Phi_m|^2} \cdot 
\left(
\begin{array}{*{4}{c}}
\Phi^0_m\Gamma_0 + i\Phi^1_m\vec\Gamma\vec\Theta& -i\Phi^1_m\Gamma_0-\Phi^0_m\vec{\Gamma}\vec\Theta& (i\Phi^3_m-\Phi^2_m)\vec{\Gamma}\vec\Theta &(-i\Phi^3_m+\Phi^2_m)\Gamma_0 \\
\Phi^0_m\vec\Gamma\vec\Theta + i\Phi^1_m\Gamma_0& -i\Phi^1_m\vec\Gamma\vec\Theta -
\Phi^0_m\Gamma_0 & (i\Phi^3_m-\Phi^2_m)\Gamma_0 &(-i\Phi^3_m+\Phi^2_m)\vec\Gamma\vec\Theta \\
(i\Phi^3_m+\Phi^2_m)\vec\Gamma\vec\Theta&-(i\Phi^3_m+\Phi^2_m)\Gamma_0 & \Phi^0_m\Gamma_0-i\Phi^1_m\vec\Gamma\vec\Theta &i\Phi^1_m\Gamma_0-\Phi^0_m\vec\Gamma\vec\Theta \\
(i\Phi^3_m+\Phi^2_m)\Gamma_0& -(i\Phi^3_m+\Phi^2_m)\vec\Gamma\vec\Theta & \Phi^0_m\vec\Gamma\vec\Theta-i\Phi^1_m\Gamma_0 & i\Phi^1_m\vec\Gamma\vec\Theta - \Phi^0_m\Gamma_0 \\
\end{array}
\right)
\eeq
with the abbreviations $\Gamma_\mu\equiv\Gamma_\mu(\Delta n)$ and 
$\Delta n = n-m$. Therefore, the first order summand of the power series 
in Eq.~(\ref{eq:DefPowerSeries}) reading
\beq
\mbox{Tr}\,\left[\DDmtwoRhoInv  B^{-1}  \right] = \sum\limits_{n}\, Tr_{8\times 8} \left[ {\cal T}_{n,n} \right] = 0
\eeq
is identical to zero and the first non-vanishing contribution is the second 
order term, which can be evaluated by explicitly computing the $8 \times 8$
trace, yielding
\bea
\label{eq:SecondOrderA4}
\mbox{Tr}\,\left[\DDmtwoRhoInv  B^{-1}  \right]^2 &=&  \sum\limits_{n_1,n_2} Tr_{8\times 8}\,\left[  {\cal T}_{n_1,n_2} {\cal T}_{n_2,n_1} \right] \nonumber \\
&=&
-8 \cdot \sum\limits_{n_1,n_2} \frac{\Phi_{n_1}^\mu \Phi_{n_2}^\mu }{|\Phi_{n_1}|^2 \cdot |\Phi_{n_2}|^2} \cdot
\left|\Gamma(\Delta n) \right|^2.
\eea
Cutting off the power series in Eq.~(\ref{eq:DefPowerSeries})
after this first non-vanishing term, which is well justified for sufficiently large values of the Yukawa
coupling constant, the effective action can be written as
\bea
\label{eq:EffectiveActionFullLargeY}
S_{eff}[\Phi] \;=\; S_\Phi &-&N_f \cdot \left(\sum\limits_{n} log(|\Phi_n|^8) +  
\frac{(4\rho)^2}{y_N^2}  \sum\limits_{n_1,n_2} \left|\Gamma(\Delta n)\right|^2\frac{\Phi_{n_1}^\dagger \Phi_{n_2}}{|\Phi_{n_1}|^2\cdot|\Phi_{n_2}|^2} 
\right) \\
&-& N_f \cdot \log\mbox{det}^*\left(B^{-1}\right) 
- N_f \cdot \log\mbox{det}^*\left(\ID + \frac{2\rho}{y_N} F[\Phi] \right) \nonumber
\eea
where the matrix $F[\Phi]$ has been defined in Eq.~(\ref{eq:DefOfFiniteTermMat}).

Some remarks concerning the remaining determinants in the latter result are in order here for the 
orientation of the reader. Here $\mbox{det}^*$ denotes the determinant over the sub-space $V_\pi$,
which has the dimension $120$. In contrast to all other terms appearing in the effective action
these determinants are {\it not} proportional to $L^4$. They are therefore suppressed as the lattice
size $L$ goes to infinity. Moreover, the very last term in Eq.~(\ref{eq:EffectiveActionFullLargeY})
even vanishes on finite lattices when the Yukawa coupling constant $y_N$ becomes large. This is 
in contrast to the determinant $\mbox{det}^*(B^{-1})$ being independent of $y_N$.
However, it is nevertheless quite instructive to consider these finite
volume effects in more detail. This can at least be done for the first determinant $\mbox{det}^*(B^{-1})$,
which can be exactly evaluated for the ansatz given in Eq.~(\ref{eq:staggeredAnsatz}) taking only a constant 
and a staggered mode of the Higgs field into account. In that case the inverse of $B$ can also be 
described in terms of a constant and a staggered mode according to
\beq
\label{eq:InverseBForstaggeredAnsatz}
\Phi_n/|\Phi_n|^2 = \hat\Phi \cdot N_f^{-\frac{1}{2}} \cdot \left(\tilde m + \tilde s\cdot (-1)^{\sum\limits_{\mu}n_\mu}
 \right), \quad
\tilde m = \frac{m}{m^2 - s^2}, \quad
\tilde s = \frac{s}{s^2 - m^2} 
\eeq
which allows to determine the desired determinant in a similar manner as 
described in Section~\ref{sec:SmallYukawaCouplings} yielding
\beq
\label{eq:DetStarBInv}
\log\mbox{det}^* \left(B^{-1}\right) = 
-60\log\left( N_f \right)
+8\log\left| \tilde m \right|
+56\log\left| \tilde m^2 - \tilde s^2 \right|.
\eeq

The obvious asymmetry in $m$ and $s$ is caused by the fact that the $8$ zero modes $\Psi^{0,\zeta\epsilon k}$ 
are not included in the sub-space $V_\pi$. The effect of the latter terms and especially the asymmetry in $m$
and $s$ is clearly observed in corresponding Monte-Carlo simulations~\cite{xypd1} on small lattices and
large values of the Yukawa coupling constant $y_N$. Moreover, the result in Eq.~(\ref{eq:DetStarBInv})
would also hinder the expectation value of the Higgs field from vanishing, thus obscuring the potential existence
of symmetric phases at large $y_N$ on small lattices. However, as the lattice size increases these finite 
volume effects eventually disappear. In the following we will therefore
neglect the $\mbox{det}^*$ terms in the effective action~(\ref{eq:EffectiveActionFullLargeY}), which is
well justified on sufficiently large lattices.

To establish the announced connection to a sigma-model we now consider the
large $N_f$-limit where the coupling constants scale according to
\beq
y_N = \tilde y_N,\, \tilde y_N = \mbox{const}, \quad
\lambda_N = \frac{\tilde \lambda_N}{N_f},\, \tilde \lambda_N = \mbox{const}, \quad
\kappa_N = \frac{\tilde \kappa_N}{N_f},\, \tilde \kappa_N = \mbox{const},
\eeq
and for the Higgs field we consider an ansatz in which the amplitude
of the local vectors $\Phi_n$ is fixed to $\varphi\in\re$ according to
\beq
\Phi_n = \sqrt{N_f} \cdot \varphi \cdot \sigma_n, \quad |\sigma_n| = 1
\eeq
where the four-component, space-time position dependent unit vectors $\sigma_n$ 
are arbitrary. In this setting the contributions to the (reduced) effective 
action are either of order $O(N_f)$ or $O(1)$. Considering only the leading 
order terms, for which the fermion doublet number $N_f$ can be completely 
factorized out, then allows to fix the Higgs field amplitude $\varphi$ by 
the determination equation 
\bea
\label{eq:FixationOfPhiBetrag}
0 &=& -4 \cdot \frac{1}{\varphi^2} + 1 + 2\tilde\lambda_N\cdot \left(\varphi^2-1
\right).
\eea
With this fixation of the Higgs field amplitude the model in Eq.~(\ref{eq:EffectiveActionFullLargeY}) 
becomes effectively a non-local, four-dimensional, non-linear sigma-model in
the large $N_f$-limit given by
\beq
\label{eq:DefOfSigmaModel}
S_{eff} = -\sum\limits_{n_1,n_2} \kappa^{eff}_{n_1,n_2} \cdot \sigma^\dagger_{n_1} \sigma_{n_2}
\eeq
with the effective, non-local coupling matrix
\beq	
\label{eq:EffectiveCouplingMatrix}
\kappa^{eff}_{n_1,n_2} =  \frac{16\rho^2}{\tilde y_N^2 \varphi^2}\left| \Gamma(\Delta n)  \right|^2 + 
\tilde\kappa_N\cdot \varphi^2\cdot\sum\limits_{\mu=\pm 1}^{\pm 4} \delta_{\Delta
n, \hat e_\mu}. 
\eeq
Here the notation "non-local" simply refers to the fact, that the field $\sigma_n$ 
at any lattice site $n$ couples itself to any other site of the lattice. 
This leaves nevertheless open the possibility that the interaction is local 
in a field theoretical sense with exponentially decaying coupling strength 
\cite{Hernandez:1998et}. We did, however, not investigate the question in 
this paper, since eventually we are mostly interested in the small Yukawa 
coupling region.

Basically, the outcome in Eq.~(\ref{eq:EffectiveCouplingMatrix}) reproduces the result which 
was found for a Higgs-Yukawa model based on Wilson fermions~\cite{Hasenfratz:1992xs} with the 
only difference that the coupling matrix in that case consisted only of nearest-neighbour 
couplings.

The phase diagram of the obtained sigma-model~(\ref{eq:DefOfSigmaModel}) can
be determined analytically by an additional large $N$-limit with $N$ denoting 
here the number of components of the vectors $\sigma_n$. The first 
step towards this evaluation is to remove the restriction $|\sigma_n|=1$ by 
introducing an auxiliary one-component, real field $\lambda_n$.
This can be done at least in two ways. One can either encode the restriction $|\sigma_n|=1$
as a $\delta$-function~\cite{Flyvbjerg:1988em} written in terms of an integration of 
$\exp(i\lambda_n(|\sigma_n|^2-1))$ over $\lambda_n$, or alternatively, one can address this 
restriction by introducing the field variables $\lambda_n$ as Lagrange-multipliers~\cite{ZinnJustin}. 
Here we follow the latter approach which leads us to the extended action
\beq
\label{eq:ActionFunctionSigmaModelLambda}
S[\sigma,\lambda] = \frac{1}{t_N}\cdot\left\{\sum\limits_{n_1,n_2}\sum\limits_{i=1}^N -\kappa^{eff}_{n_1,n_2} \cdot
\sigma^i_{n_1}\cdot \sigma^i_{n_2} + \sum\limits_{n}\lambda_n
\cdot\left( \sum\limits_{i=1}^N\left[\sigma^i_n\right]^2 -1\right)\right\}
\eeq
the minima of which can now be searched for without having to consider 
any restriction on the Higgs field amplitude. Here, an additional parameter $t_N$ was introduced. 
For $t_N=1$ the given action corresponds to the prior form of the action. This new parameter is inserted, 
since it will allow to factorize a factor $N$ out of the action as required by the large $N$ approach. 
This can be achieved by scaling $t_N$ according to 
\beq
t_N = \frac{\tilde t_N}{N}, \quad \tilde t_N = \mbox{const}, 
\eeq
where we choose the setting $\tilde t_N=4$, since this recovers our actual effective sigma-model at $N=4$.

The remaining problem to solve is to find the minimum of the action $S[\sigma,\lambda]$. However, it is well 
known from investigations of pure sigma-models that the phase transitions of such models cannot be correctly
determined by evaluating the effective action $S[\sigma,\lambda]$ in Eq.~(\ref{eq:ActionFunctionSigmaModelLambda})
directly by restricting the consideration to only some selected modes of the fields $\sigma$ and $\lambda$. 
(Doing so would yield a first order phase transition at $\tilde \kappa_N=0$.)
This is in contrast to the situation we discussed in Section~\ref{sec:SmallYukawaCouplings}.
Instead, we first integrate out all modes of all $N$ components of the 
field $\sigma$ except for the constant and staggered modes. This can be done by taking only the constant
mode of the {\it auxiliary field} $\lambda_n$ into account, \ie $\lambda_n\equiv\LagrangeMul=\mbox{const}$. 
Doing so reduces the action $S[\sigma,\lambda]$ to
\bea
\label{eq:EffectiveActionReducedModel}
S[m^i,s^i,\LagrangeMul] &=& -\ln\left[ \mbox{det}'\left( -\kappa^{eff} + \LagrangeMul   \right)   \right]^{-N/2} 
+ \frac{1}{t_N} \Bigg\{ \sum\limits_{i=1}^N \left[m^i\right]^2 \cdot \left\langle0\left| -\kappa^{eff} + \LagrangeMul
\right|0\right\rangle \nonumber \\
&+& \sum\limits_{i=1}^N \left[s^i\right]^2 \cdot \left\langle\pi\left| -\kappa^{eff} + \LagrangeMul \right|\pi\right\rangle
-L^4 \LagrangeMul  \Bigg\},
\eea
depending only on the real scalar $\LagrangeMul$ and the amplitudes $m^i$, $s^i$ of the constant and
staggered modes, respectively. Here the notations $|0\rangle$ and $|\pi\rangle$ were used, denoting the 
constant and staggered modes (normalized by a factor $1/\sqrt{L^4}$) according to
\beq
\label{eq:DefOfBRACKETmodes}
|k\rangle \; \equiv \; \sqrt{\frac{1}{L^4}} \; e^{ik\cdot n}
\eeq
being eigenvectors of $\kappa^{eff}$ and $\mbox{det}'$ is the determinant neglecting the two latter modes.
For convenience, the introduced short-hand notation $0 \equiv (0,0,0,0)$ and $\pi\equiv (\pi,\pi,\pi,\pi)$ will 
also be applied in the following where it is unambiguous.

One remark is in order here for the orientation of the reader. The performed
Gauss-integrations are only well-defined, if the involved eigenvalues 
of the operator $-\kappa^{eff} + \LagrangeMul$ are positive, which is not guaranteed 
at this point. However, this step will be justified (and made more precise) {\it a posteriori} 
when a certain value for the scalar $\LagrangeMul$ will be assumed by solving the resulting 
gap equations. Here we will first continue with this formal expression and postpone its 
further discussion to the end of this section.

To evaluate this latter determinant, the eigenvalues of the matrix $\kappa^{eff}$ need to be known.
The eigenvectors are simply plane waves with wave vectors $k\in\ImpSpace$ and one easily finds the corresponding 
eigenvalues according to
\bea
\label{eq:EigenValuesOfCouplingMatrixP}
\sum\limits_{n_2}\kappa^{eff}_{n_1,n_2} \cdot e^{ikn_2} 
&=& 
\left(   2\tilde\kappa_N \varphi^2 \sum\limits_{\mu= 1}^{4}\cos(k_\mu) 
+ \frac{16\rho^2}{\tilde y_N^2 \varphi^2}\cdot q(k)\right) \cdot e^{ikn_1}
\eea
where $q(k)$ denotes the eigenvalues of the matrix $|\Gamma(\Delta n)|^2$ given by
\beq
\re\ni q(k) = \frac{1}{L^4}\sum\limits_{p\in\ImpSpace} 
\alpha^+(p)\cdot \alpha^+(\wp)\cdot \frac{\tilde p\cdot \tilde \wp}{\sqrt{\tilde p^2}\cdot
\sqrt{\tilde{\wp}^2}}, \quad \wp = k-p.
\eeq
For the numerical evaluation of this quantity it is useful to use some symmetries of $q(k)$. One 
has $q(k) = q(k')$ at least, if $k'$ is a permutation of the components of $k$, or
if $k'_\mu = \pm k_\mu$ for all $\mu$.

Now we can search for the absolute minima of the effective action in 
Eq.~(\ref{eq:EffectiveActionReducedModel}). For this purpose we relate the
amplitudes $m^i$, $s^i$ to the values of the overall magnetization $m$ 
and staggered magnetization $s$, respectively, according to
\beq
\label{eq:RelOfMagToMode}
m^i = \sqrt{\frac{L^4}{N}}\, m \quad \mbox{ and } \quad s^i = \sqrt{\frac{L^4}{N}}\, s. 
\eeq
With this notation one directly obtains from the effective action in Eq.~(\ref{eq:EffectiveActionReducedModel}) 
the following expression in terms of the quantities $m,s$ and $\LagrangeMul$
\bea
\label{eq:AlternativeApproachEffectiveActionSigmaModel}
S[m,s,\LagrangeMul] &=& \frac{N}{2}\mbox{Tr}'\,\ln\left[ -\kappa^{eff} + \LagrangeMul \right] 
+ \frac{N}{\tilde t_N}  \cdot m^2 \cdot L^4 \cdot \left( -8\tilde\kappa_N\varphi^2 - \frac{16\rho^2}{\tilde y_N^2\varphi^2}q(0) + \LagrangeMul  \right) \nonumber \\
&+& \frac{N}{\tilde t_N}  \cdot s^2 \cdot L^4 \cdot \left( +8\tilde\kappa_N\varphi^2 - \frac{16\rho^2}{\tilde y_N^2\varphi^2}q(\pi) + \LagrangeMul  \right)
-\frac{N}{\tilde t_N}L^4 \LagrangeMul,
\eea
where the summation over the coupling matrix components has been performed 
by using Eq.~(\ref{eq:EigenValuesOfCouplingMatrixP}) with the settings 
$k=(0,0,0,0)\equiv 0$ 
and $k=(\pi,\pi,\pi,\pi)\equiv \pi$, respectively. Analogously to $\mbox{det}'$, 
$\mbox{Tr}'$ denotes the trace neglecting the modes $|0\rangle$ and 
$|\pi\rangle$. We can now derive the corresponding gap equations by 
differentiating with respect to $m$, $s$, and $\LagrangeMul$ leading to

\beq
\label{eq:LargeYGap1}
0=m\cdot \left[\LagrangeMul - \left(8\tilde\kappa_N \varphi^2 + \frac{16\rho^2}{\tilde y_N^2 \varphi^2} \cdot q(0)\right) 
\right],
\eeq

\beq
\label{eq:LargeYGap2}
0=s\cdot \left[\LagrangeMul - \left(-8\tilde\kappa_N \varphi^2 + \frac{16\rho^2}{\tilde y_N^2 \varphi^2} \cdot  q\left(
\pi\right)  \right)  \right],
\eeq

\beq
\label{eq:LargeYGap3}
m^2+s^2 =  1 - \frac{\tilde t_N}{4} \frac{1}{L^4}\sum\limits_{{k\in\ImpSpace}\atop{0\neq k \neq \pi}} 
\left[-\tilde\kappa_N \varphi^2 \sum\limits_{\mu= 1}^{ 4}\cos(k_\mu) - \frac{8\rho^2}{\tilde y_N^2 \varphi^2} q(k) +
\frac{\LagrangeMul}{2}\right]^{-1}.
\eeq

\bc
\setlength{\unitlength}{0.01mm}
\begin{figure}[htb]
\begin{tabular}{cc}
$\tilde\lambda_N=0.0$ & $\tilde\lambda_N=0.1$ \\
\begin{picture}(6600,5500)
\put(600,500){\includegraphics[width=5cm]{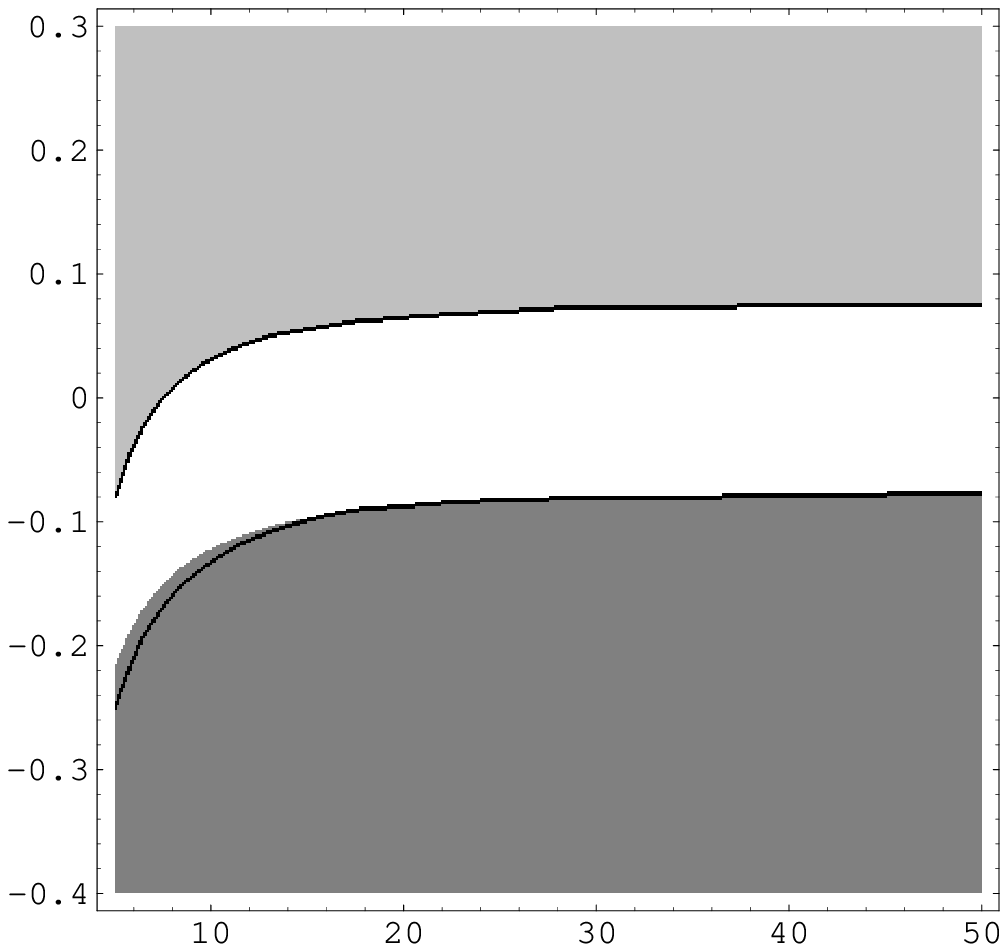}}
\put(25,3100){$\tilde \kappa_N$}
\put(3200,300){$\tilde y_N$}
\put(4500,3250){\textbf {SYM}}
\put(1500,4500){\textbf {FM}}
\put(4500,1100){\textbf {AFM}}
\end{picture}
&
\begin{picture}(6600,5500)
\put(600,500){\includegraphics[width=5cm]{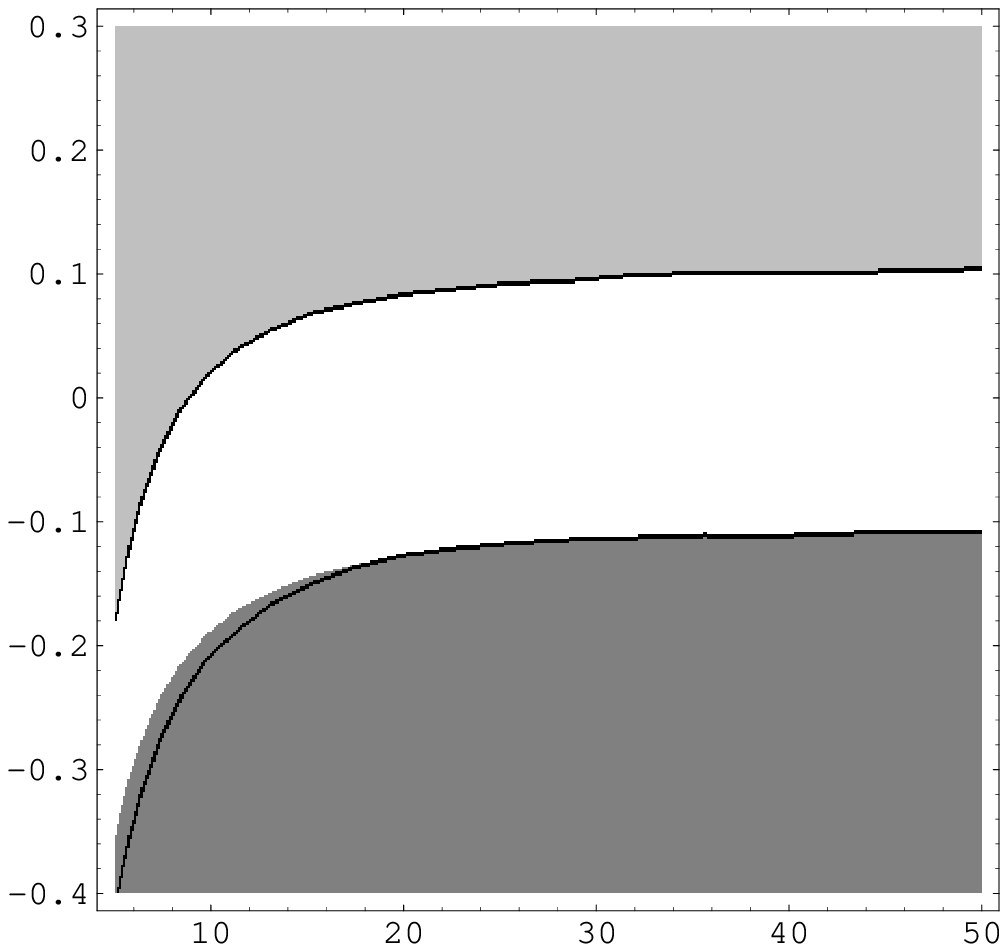}}
\put(25,3100){$\tilde \kappa_N$}
\put(3200,300){$\tilde y_N$}
\put(4500,3250){\textbf {SYM}}
\put(1500,4500){\textbf {FM}}
\put(4500,1100){\textbf {AFM}}
\end{picture}
\\
$\tilde\lambda_N=1.0$ & $\tilde\lambda_N=10.0$ \\
\begin{picture}(6600,5500)
\put(600,500){\includegraphics[width=5cm]{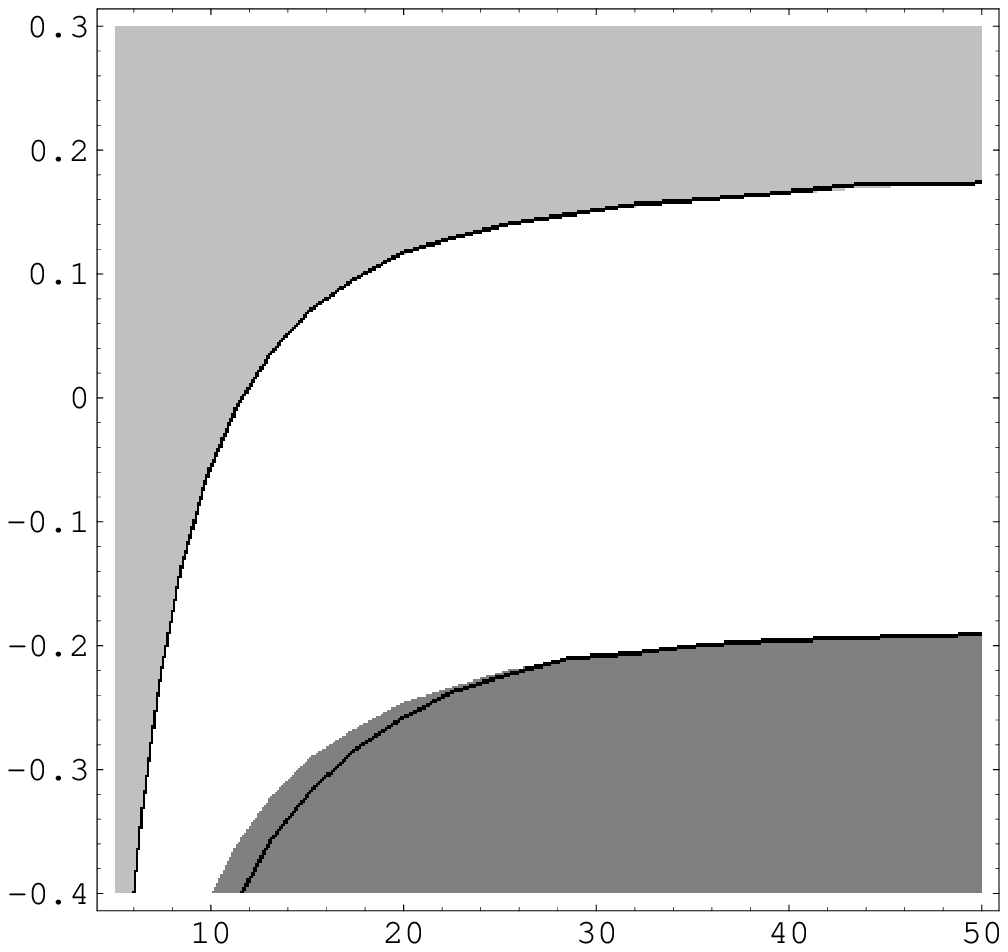}}
\put(25,3100){$\tilde \kappa_N$}
\put(3200,300){$\tilde y_N$}
\put(4500,3250){\textbf {SYM}}
\put(1500,4500){\textbf {FM}}
\put(4500,1100){\textbf {AFM}}
\end{picture}
&
\begin{picture}(6600,5500)
\put(600,500){\includegraphics[width=5cm]{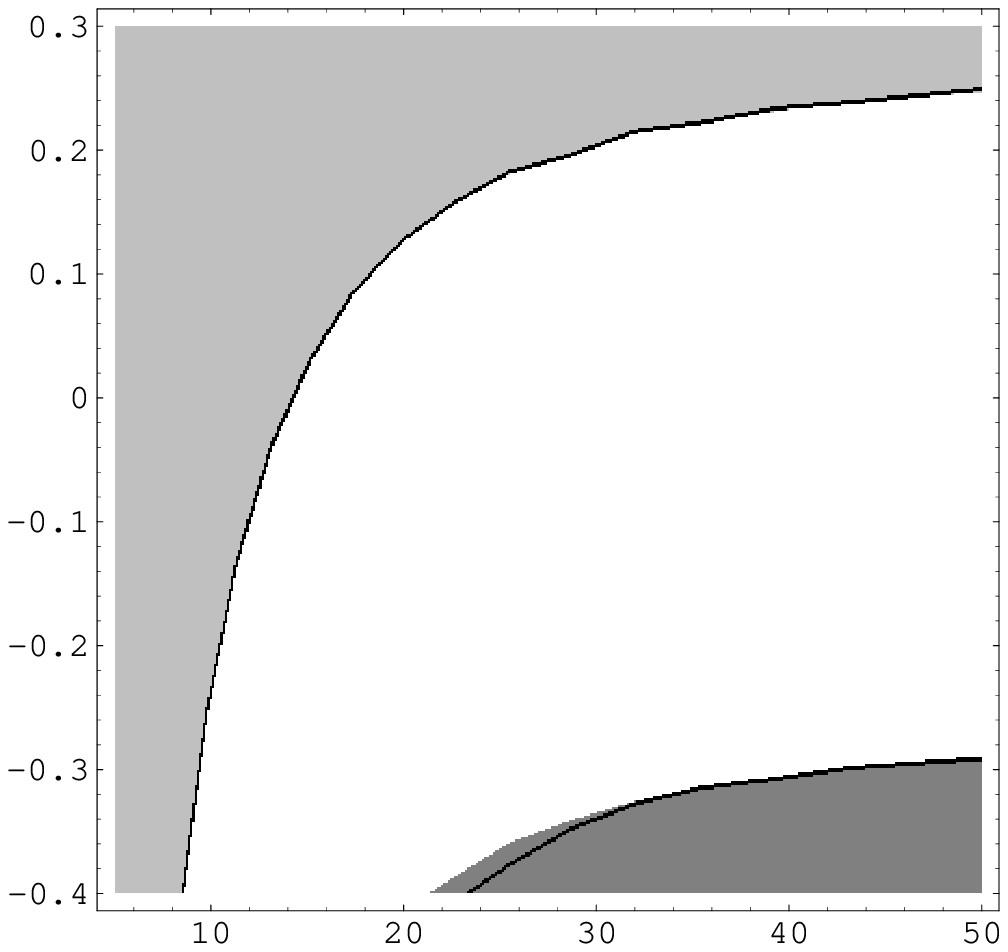}}
\put(25,3100){$\tilde \kappa_N$}
\put(3200,300){$\tilde y_N$}
\put(4500,3250){\textbf {SYM}}
\put(1500,4500){\textbf {FM}}
\put(4500,1100){\textbf {AFM}}
\end{picture}
\end{tabular}
\caption{Phase diagrams for $L=\infty$ with respect to the Yukawa coupling constant $\tilde y_N$ and the hopping 
parameter $\tilde \kappa_N$ for several selected values of the quartic coupling constant $\tilde \lambda_N$.
The presented phase structure was determined for $\epsilon=10^{-1}$, while the black lines show
the phase transition lines obtained for $\epsilon=10^{-3}$. An explanation of the $\epsilon$-dependence
of the presented results is given in the text.}
\label{fig:PhaseDiagrams2}
\end{figure}
\ec

Equation~(\ref{eq:LargeYGap1}) implies that $m$ or the given argument within the square brackets has to
vanish. An analogous observation can be drawn from Eq.~(\ref{eq:LargeYGap2}). For the investigation
of the phase structure we now consider two different scenarios for the amplitudes $m$ and $s$, namely
a ferromagnetic phase ($m\neq 0$, $s=0$) and an anti-ferromagnetic phase ($m= 0$, $s\neq0$). For each
of these cases we can then derive a self-consistency relation:

\begin{enumerate}
\item For a ferromagnetic phase~(FM) ($m\neq 0$, $s=0$) one obtains from (\ref{eq:LargeYGap1})
\beq
\label{eq:LargeYGap4}
\LagrangeMul = 8\tilde\kappa_N \varphi^2 + \frac{16\rho^2}{\tilde y_N^2 \varphi^2} \cdot q(0)
\eeq
and hence the following self-consistency relation
\beq
\label{eq:LargeYGap5}
0<m^2 =  1 - \frac{\tilde t_N}{4}\frac{1}{L^4}\sum\limits_{{k\in\ImpSpace_m(\epsilon)}\atop{0\neq k \neq \pi}} 
\Bigg[\underbrace{\tilde\kappa_N \varphi^2 \sum\limits_{\mu= 1}^{ 4}\left(1-\cos(k_\mu)\right) 
+ \frac{8\rho^2}{\tilde y_N^2 \varphi^2} \left(q(0)-q(k)\right)}_{{\cal W}_m(k)}\Bigg]^{-1}.
\eeq
\item For an anti-ferromagnetic phase~(AFM) ($m=0$, $s\neq 0$) one obtains from (\ref{eq:LargeYGap2})
\beq
\label{eq:LargeYGap6}
\LagrangeMul = -8\tilde\kappa_N \varphi^2 + \frac{16\rho^2}{\tilde y_N^2 \varphi^2} \cdot q\left( \pi\right)
\eeq
and hence the self-consistency relation
\beq
\label{eq:LargeYGap7}
0<s^2 =  1 - \frac{\tilde t_N}{4} \frac{1}{L^4}\sum\limits_{{k\in\ImpSpace_s(\epsilon)}\atop{0\neq k \neq \pi}} 
\Bigg[\underbrace{-\tilde\kappa_N \varphi^2 \sum\limits_{\mu= 1}^{ 4}\left(1+\cos(k_\mu)\right) 
+ \frac{8\rho^2}{\tilde y_N^2 \varphi^2} \left(q\left( \pi\right)-q(k)\right)}_{{\cal W}_s(k)}\Bigg]^{-1}.
\eeq
\end{enumerate}

\bc
\setlength{\unitlength}{0.01mm}
\begin{figure}[htb]
\begin{picture}(12000,9060)
\put(0,1000){\includegraphics[angle=0,width=0.2676\textwidth]{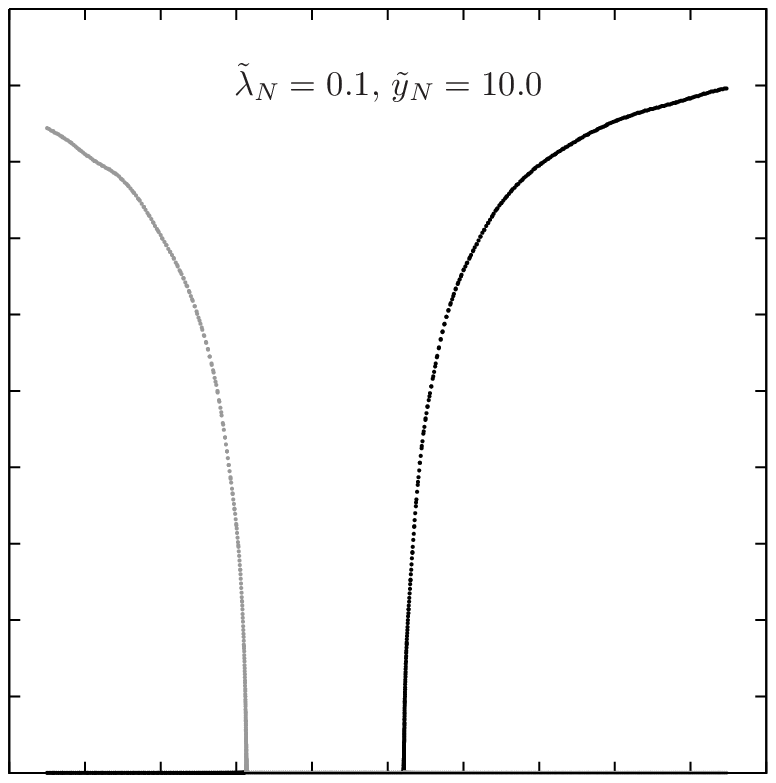}}
\put(4000,1000){\includegraphics[angle=0,width=0.2676\textwidth]{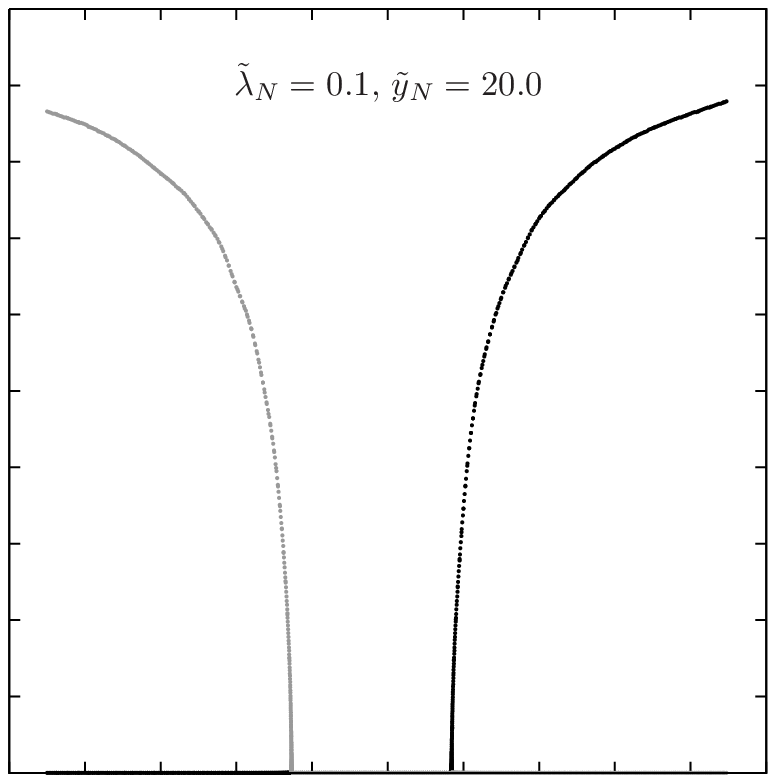}}
\put(8000,1000){\includegraphics[angle=0,width=0.2676\textwidth]{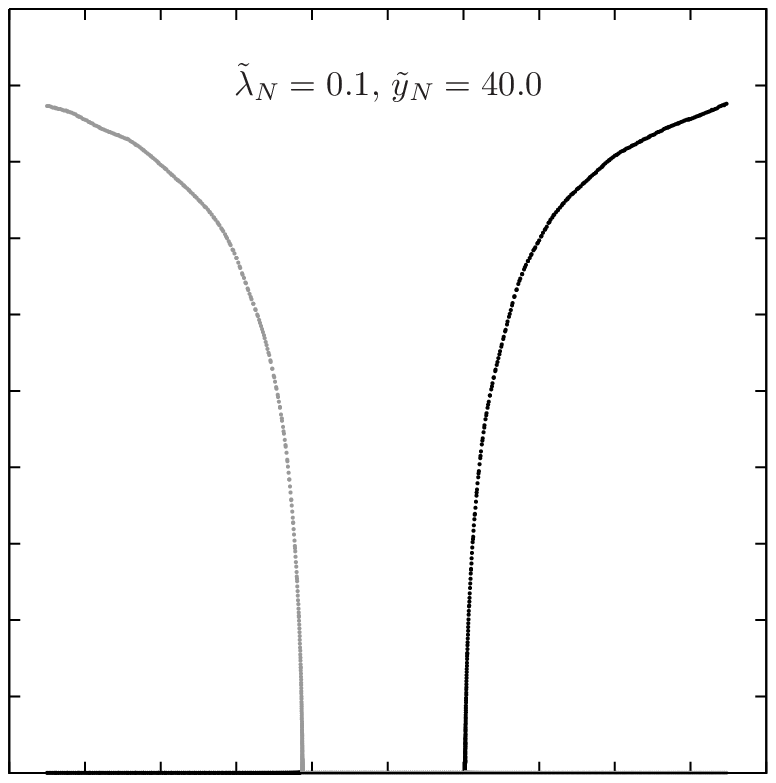}}
\put(0,5030){\includegraphics[angle=0,width=0.2676\textwidth]{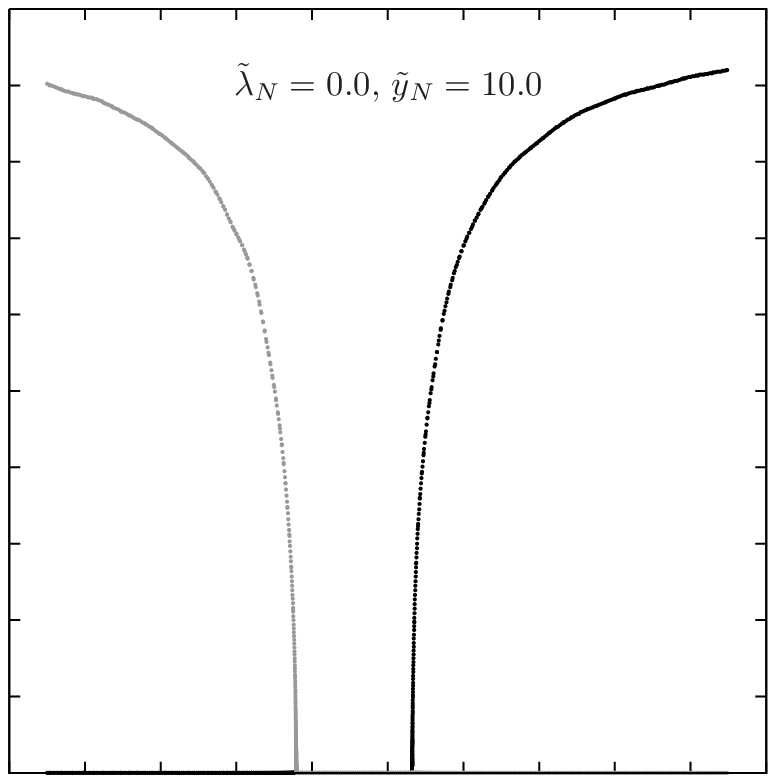}}
\put(4000,5030){\includegraphics[angle=0,width=0.2676\textwidth]{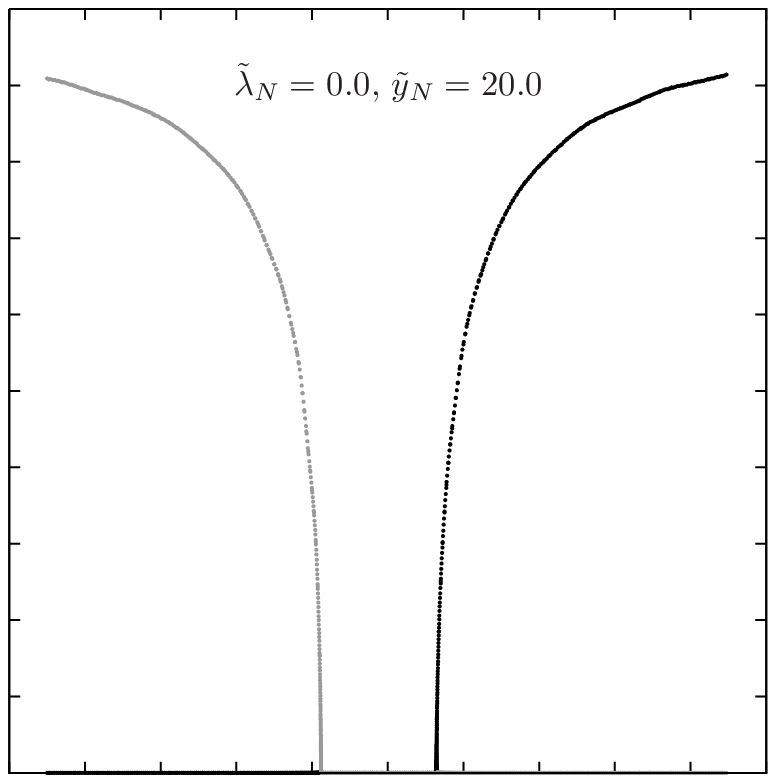}}
\put(8000,5030){\includegraphics[angle=0,width=0.2676\textwidth]{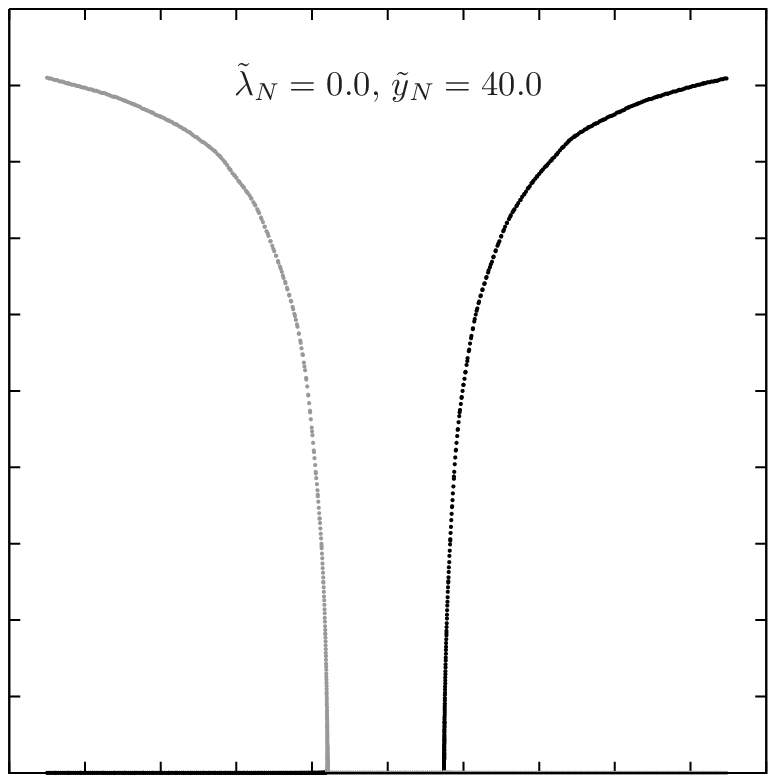}}

\put(+178,700){\tiny{-0.4}}
\put(+978,700){\tiny{-0.2}}
\put(1855,700){\tiny{0.0}}
\put(2655,700){\tiny{0.2}}
\put(3455,700){\tiny{0.4}}

\put(4178,700){\tiny{-0.4}}
\put(4978,700){\tiny{-0.2}}
\put(5855,700){\tiny{0.0}}
\put(6655,700){\tiny{0.2}}
\put(7455,700){\tiny{0.4}}

\put(8178,700){\tiny{-0.4}}
\put(8978,700){\tiny{-0.2}}
\put(9855,700){\tiny{0.0}}
\put(10655,700){\tiny{0.2}}
\put(11455,700){\tiny{0.4}}

\put(5800,100){$\tilde \kappa_N$}

\put(-430,960){\tiny{0.0}}
\put(-430,1766){\tiny{0.2}}
\put(-430,2572){\tiny{0.4}}
\put(-430,3378){\tiny{0.6}}
\put(-430,4189){\tiny{0.8}}

\put(-430,4990){\tiny{0.0}}
\put(-430,5796){\tiny{0.2}}
\put(-430,6602){\tiny{0.4}}
\put(-430,7408){\tiny{0.6}}
\put(-430,8219){\tiny{0.8}}
\put(-430,9020){\tiny{1.0}}
\end{picture}
\caption{Expectation values for the amplitudes of the constant ($m$: black curve) and staggered ($s$: gray curve) modes 
for several selected values of the Yukawa coupling constant $\tilde y_N$ and the quartic coupling parameters $\tilde \lambda_N=0.0$
and $\tilde \lambda_N=0.1$.
The results were obtained for $L=\infty$.}
\label{fig:ExpectedMSPlots3}
\end{figure}
\ec

Three further remarks shall be given here.

{\it (I)} The equations~(\ref{eq:LargeYGap5}) and~(\ref{eq:LargeYGap7}) are denoted 
as self-consistency relations because the assumption of a (anti-)ferromagnetic phase becomes 
inconsistent, if the resulting value for $m^2$ (or $s^2$, respectively) becomes non-positive. 
If both assumptions become inconsistent simultaneously, this corresponds to a symmetric 
phase~(SYM) with $m=s=0$, while the case $m^2>0$ and $s^2>0$ is denoted as a ferrimagnetic 
phase~(FI).

{\it (II)} For the ferromagnetic phase the choice of $\LagrangeMul$ according to Eq.~(\ref{eq:LargeYGap4})
justifies the integration performed in Eq.~(\ref{eq:EffectiveActionReducedModel}) {\it a posteriori}, 
because it sufficiently shifts the eigenvalues $2{\cal W}_m(k)$ of the matrix $-\kappa^{eff}+\LagrangeMul$ to 
make all of them positive, except for the constant mode ($k=0$) which was excluded from the Gauss-integration.

{\it (III)} For the anti-ferromagnetic phase, in contrast, choosing $\LagrangeMul$ according to Eq.~(\ref{eq:LargeYGap6})
does not guarantee all eigenvalues $2{\cal W}_s(k)$ of $-\kappa^{eff}+\LagrangeMul$ to be positive. The Gauss-integration in 
Eq.~(\ref{eq:EffectiveActionReducedModel}) can therefore only be performed for all those modes $0\neq k\neq \pi$ 
which fulfill ${\cal W}_s(k)\ge\epsilon$ with an arbitrary lower bound $\epsilon>0$. The details of this statement 
are presented in Appendix~\ref{AppendixEpsilonCut}.
The results of this more careful consideration are already presented in Eq.~(\ref{eq:LargeYGap5}) and 
Eq.~(\ref{eq:LargeYGap7}). The only difference to the naive result is that the set over which the sum has to be
performed is reduced from $\ImpSpace$ to $\ImpSpace_s(\epsilon)$  with the definitions
\bea
\label{eq:DefOfRedImpSpaceMS}
\ImpSpace_m(\epsilon) = \Big\{k\in\ImpSpace\,:\, {\cal W}_m(k)\ge\epsilon  \Big\} & \mbox{ and } &
\ImpSpace_s(\epsilon) = \Big\{k\in\ImpSpace\,:\, {\cal W}_s(k)\ge\epsilon  \Big\},
\eea
where the introduction of the set $\ImpSpace_m(\epsilon)$ is actually unnecessary due to the
previous remark {\it (II)}.

\bc
\setlength{\unitlength}{0.01mm}
\begin{figure}[htb]
\begin{picture}(12000,9060)
\put(0,1000){\includegraphics[angle=0,width=0.2676\textwidth]{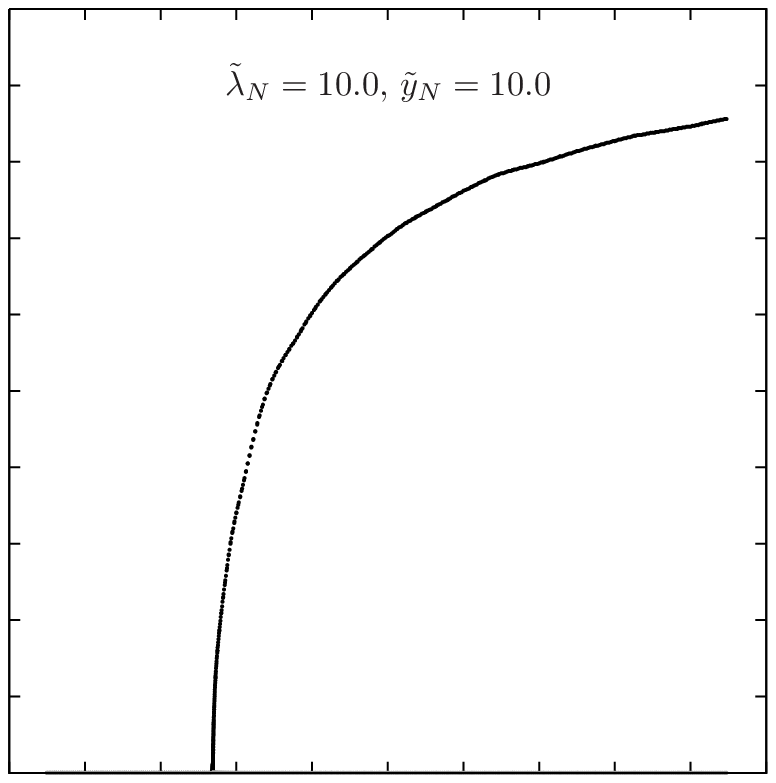}}
\put(4000,1000){\includegraphics[angle=0,width=0.2676\textwidth]{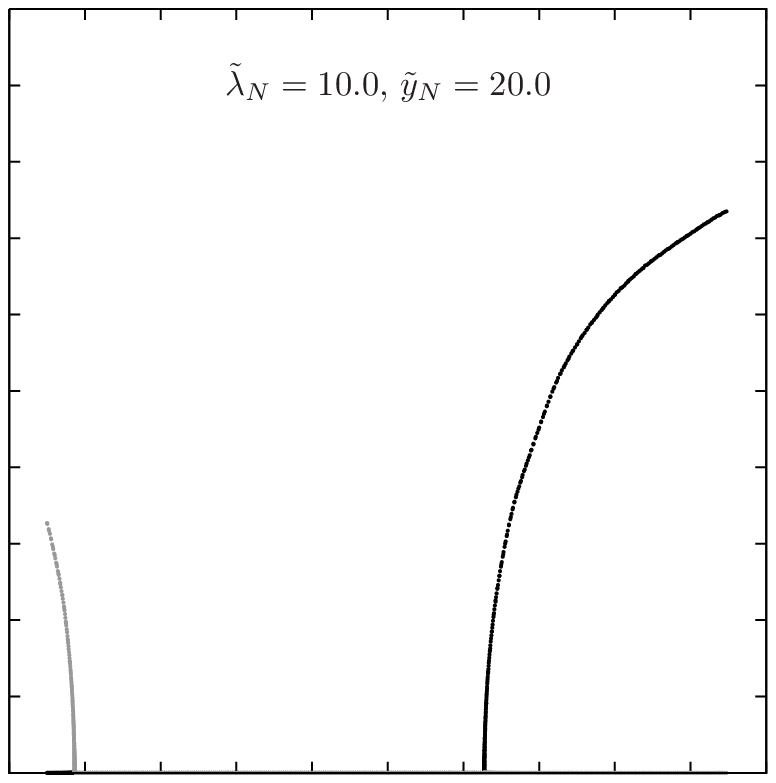}}
\put(8000,1000){\includegraphics[angle=0,width=0.2676\textwidth]{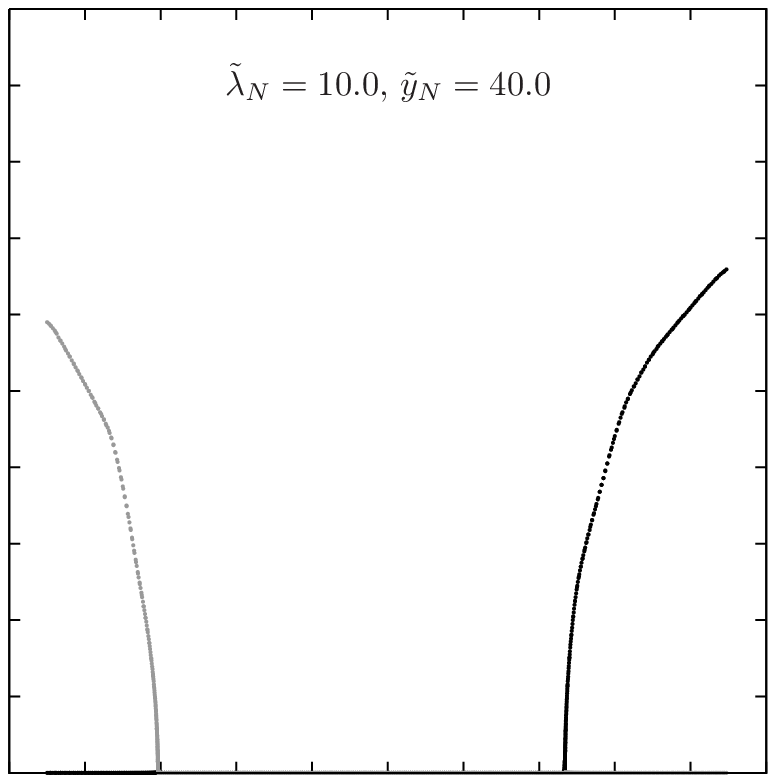}}
\put(0,5030){\includegraphics[angle=0,width=0.2676\textwidth]{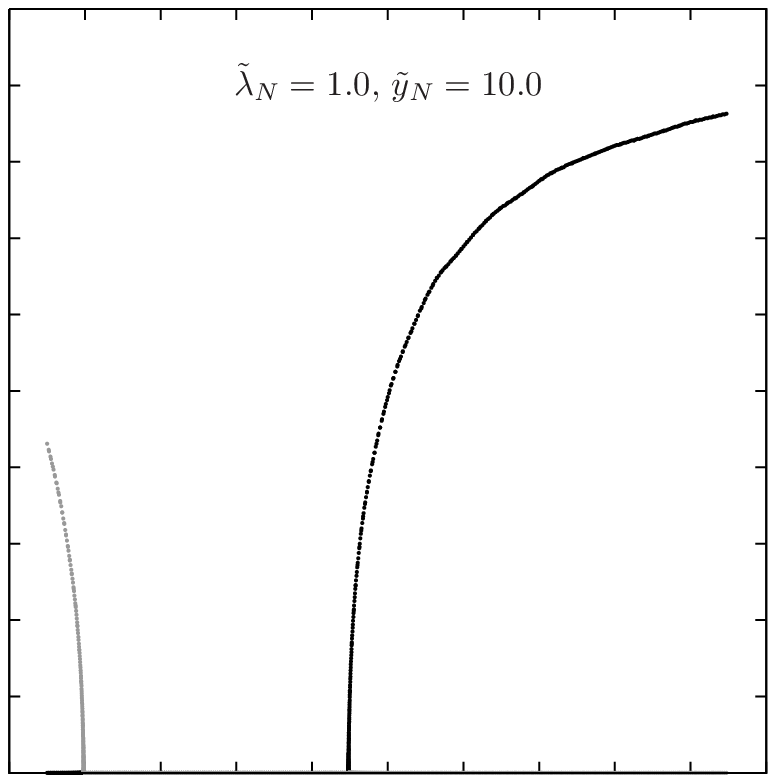}}
\put(4000,5030){\includegraphics[angle=0,width=0.2676\textwidth]{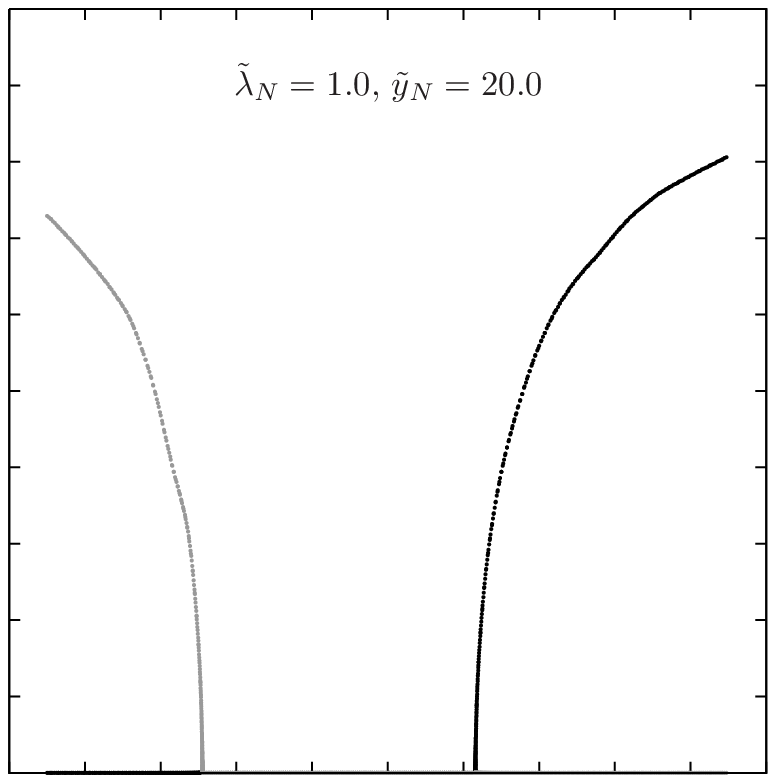}}
\put(8000,5030){\includegraphics[angle=0,width=0.2676\textwidth]{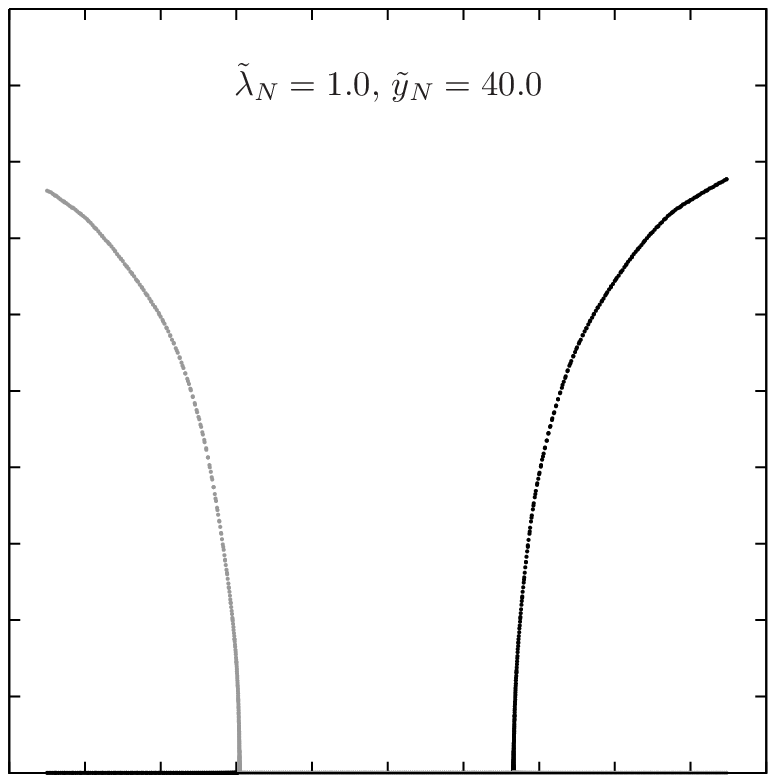}}

\put(+178,700){\tiny{-0.4}}
\put(+978,700){\tiny{-0.2}}
\put(1855,700){\tiny{0.0}}
\put(2655,700){\tiny{0.2}}
\put(3455,700){\tiny{0.4}}

\put(4178,700){\tiny{-0.4}}
\put(4978,700){\tiny{-0.2}}
\put(5855,700){\tiny{0.0}}
\put(6655,700){\tiny{0.2}}
\put(7455,700){\tiny{0.4}}

\put(8178,700){\tiny{-0.4}}
\put(8978,700){\tiny{-0.2}}
\put(9855,700){\tiny{0.0}}
\put(10655,700){\tiny{0.2}}
\put(11455,700){\tiny{0.4}}

\put(5800,100){$\tilde \kappa_N$}

\put(-430,960){\tiny{0.0}}
\put(-430,1766){\tiny{0.2}}
\put(-430,2572){\tiny{0.4}}
\put(-430,3378){\tiny{0.6}}
\put(-430,4189){\tiny{0.8}}

\put(-430,4990){\tiny{0.0}}
\put(-430,5796){\tiny{0.2}}
\put(-430,6602){\tiny{0.4}}
\put(-430,7408){\tiny{0.6}}
\put(-430,8219){\tiny{0.8}}
\put(-430,9020){\tiny{1.0}}
\end{picture}
\caption{Expectation values for the amplitudes of the constant ($m$: black curve) and staggered ($s$: gray curve) modes 
for several selected values of the Yukawa coupling constant $\tilde y_N$ and the quartic coupling parameters $\tilde \lambda_N=1.0$
and $\tilde \lambda_N=10.0$.
The results were obtained for $L=\infty$.}
\label{fig:ExpectedMSPlots4}
\end{figure}
\ec

The corresponding phase structure can now be obtained by numerically evaluating equations~(\ref{eq:LargeYGap5}) 
and~(\ref{eq:LargeYGap7}). For some selected values of the quartic coupling $\tilde \lambda_N$ the resulting
phase diagrams with respect to the parameters $\tilde\kappa_N$ and $\tilde y_N$ are shown in Fig.~\ref{fig:PhaseDiagrams2}.
All presented results were obtained for an infinite lattice, \ie $L=\infty$.
For $\tilde y_N\rightarrow\infty$ the effective coupling matrix in Eq.~(\ref{eq:EffectiveCouplingMatrix}) converges
to the coupling structure of a pure nearest-neighbour sigma-model. One therefore expects a symmetric phase centered
around $\tilde \kappa_N=0$ at large values of the Yukawa coupling constant $\tilde y_N$ as can be observed in the
plots. For decreasing $\tilde y_N$ the symmetric phase bends towards negative values of $\tilde \kappa_N$. In the plots 
the results for the phase transition lines obtained for $\epsilon=10^{-1}$ and $\epsilon=10^{-3}$ are compared to each other. 
While the phase transition line to the ferromagnetic phase is unaffected by small changes to $\epsilon$ as expected, 
the curves start to differ for the anti-ferromagnetic phase transition at small values of $\tilde y_N$. The discrepancy 
between these two lines can serve as an indicator down to which value of $\tilde y_N$ the neglection of the modes 
with ${\cal W}_s(k)<\epsilon$ can be considered as a good approximation (besides the uncertainties arising from cutting off
the power series in Eq.~(\ref{eq:DefPowerSeries}) at small values of $\tilde y_N$). 
We add here, that we chose the presented parameter range in all phase diagrams
such that the volume of the space of the considered modes is at least 
$95\%$ of the volume of the whole mode space, \ie
$\mbox{Vol}(\ImpSpace_s(\epsilon))\ge 0.95\cdot \mbox{Vol}(\ImpSpace)$. For $\tilde y_N\rightarrow\infty$ the volume
of the neglected modes vanishes and the problem encountered during the Gauss-integration in Eq.~(\ref{eq:EffectiveActionReducedModel})
eventually disappears.

The order of the phase transitions can again be determined by calculating the expectation values
of the amplitudes of the constant and staggered modes $m$ and $s$ directly from equations~(\ref{eq:LargeYGap5}) 
and~(\ref{eq:LargeYGap7}). The corresponding results are presented in Fig.~\ref{fig:ExpectedMSPlots3} and
Fig.~\ref{fig:ExpectedMSPlots4}. One clearly sees that the occurring phase transitions are of second order as
one would also expect from the limit $\tilde y_N\rightarrow\infty$ where the model becomes a sigma-model.

\section{Summary and conclusions}
\label{sec:Conclusion}
In this paper we have studied analytically the phase structure of a chirally invariant lattice Higgs-Yukawa 
model, originally proposed by L\"uscher. This was possible in the large $N_f$-limit for small as well as for 
large values of the Yukawa coupling constant and it could be shown that the model possesses a rich 
phase structure.

In Section~\ref{sec:SmallYukawaCouplings} we began by considering the model at small values 
of the Yukawa and quartic coupling constant
and argued that taking only the constant ($m$) and staggered ($s$) modes of the Higgs field 
into account is sufficient for the 
determination of the phases in that regime of the Yukawa and quartic coupling constant. We then presented 
an explicit expression for the effective potential at
tree-level in terms of $m$ and $s$ and showed the corresponding phase diagrams for some 
selected values of the quartic coupling
constant. In these diagrams all possible phases, \ie symmetric ($m=0$, $s=0$), ferromagnetic ($m\neq 0$, $s=0$), 
anti-ferromagnetic ($m=0$, $s\neq 0$), and ferrimagnetic phases ($m\neq 0$, $s\neq 0$), could 
be observed. Furthermore, we concluded from our result for the effective potential that 
the occurring phase transitions from the symmetric to the ferromagnetic and anti-ferromagnetic
phases are of second order.

In the following Section~\ref{sec:LargeYukawaCouplings} we proceeded to the regime of 
large values of the Yukawa coupling constant $y_N$. We showed that for sufficiently large 
values of $y_N$ and arbitrary values of the quartic coupling constant $\lambda_N$
the model becomes an $O(4)$-symmetric, non-linear sigma-model
in the large $N_f$-limit up to some finite-volume terms. In particular, this relation to a 
sigma-model has the consequence that
a symmetric phase also exists at large values of the Yukawa coupling constant. We determined the 
phase structure of the latter
sigma-model by an additional large $N$-limit with $N$ denoting the number of Higgs field 
components here. The corresponding
phase diagrams revealed again a rich structure consisting of symmetric, ferromagnetic, 
and anti-ferromagnetic phases separated 
by second order phase transitions. The symmetric phase, however, was shown to emerge only 
in the infinite volume limit. For small lattices,  
finite volume effects
cause an asymmetry in $m$ and $s$ which one would not expect in a pure sigma-model. 
These finite volume effects may easily give rise to a misleading interpretation that 
a symmetric phase at strong values of the Yukawa coupling constant does not exist.
However, on sufficiently 
large lattices the symmetric phase should become clearly observable and the asymmetry should disappear. 

The validity of our analytical results and in particular the latter predictions about the 
symmetric phase at large
$y_N$ will be confronted in an upcoming paper with the results of corresponding 
Monte-Carlo simulations including the chiral invariant fermions in a fully dynamical 
fashion.

\begin{appendix}

\section{}
\label{AppendixRthPowerTres}
In this appendix we would like to make up for the neglected derivation of
Eq.~(\ref{eq:TrGn1}). Starting from Eq.~(\ref{eq:CalcOfABInv}) one finds
\bea
\label{eq:TrGn1FULL}
Tr\, \left[ \DDmtwoRhoInv B^{-1} \right]^r &=& \sum\limits_{n_1,...,n_r} Tr_{8\times 8}\, \left(\left[ \DDmtwoRhoInv B^{-1} \right]_{n_1,n_2}\cdot ... \cdot 
\left[ \DDmtwoRhoInv B^{-1} \right]_{n_r,n_1} \right)  \\
&=& \sum\limits_{n_1,...,n_r} \sum\limits_{{\zeta_1\epsilon_1 k_1,...,\zeta_r\epsilon_r k_r}\atop{\zeta_1'\epsilon_1' k_1',...,\zeta_r'\epsilon_r' k_r'}}
\sum\limits_{p_1,...,p_r\in\ImpSpace}
\frac{e^{ip_1(n_1-n_2)}}{L^4} \cdot ... \cdot \frac{e^{ip_r(n_r-n_1)}}{L^4} \nonumber \\
&\times& Tr_{8\times 8}\, \Bigg[ 
u^{\zeta_1\epsilon_1 k_1}(p_1) \alpha^{\epsilon_1}(p_1) 
\left(\hat B^{(p_1)}(\Phi^*_{n_2}/|\Phi_{n_2}|^2)\right)_{\zeta_1\epsilon_1k_1,\zeta_1'\epsilon_1'k_1'} 
\underbrace{\left[u^{\zeta_1'\epsilon_1'k_1'}(p_1)\right]^\dagger 
\cdot u^{\zeta_2\epsilon_2 k_2}(p_2)}_{U(p_1,p_2)_{\zeta_1'\epsilon_1' k_1',\zeta_2\epsilon_2 k_2}} \nonumber\\
&\times& \alpha^{\epsilon_2}(p_2) \left(\hat B^{(p_2)}(\Phi^*_{n_3}/|\Phi_{n_3}|^2)\right)_{\zeta_2\epsilon_2k_2,\zeta_2'\epsilon_2'k_2'} 
\left[u^{\zeta_2'\epsilon_2' k_2'}(p_2)\right]^\dagger \cdot\,.\,.\,.\, \cdot
u^{\zeta_r\epsilon_r k_r}(p_r) \alpha^{\epsilon_r}(p_r) \nonumber\\
&\times& \left(\hat B^{(p_r)}(\Phi^*_{n_1}/|\Phi_{n_1}|^2)\right)_{\zeta_r\epsilon_nk_r,\zeta_r'\epsilon_r'k_r'} 
\left[u^{\zeta_r'\epsilon_r' k_r'}(p_r)\right]^\dagger \Bigg] \nonumber \\
&=& \sum\limits_{n_1,...,n_r} \sum\limits_{p_1,...,p_r\in\ImpSpace}
Tr_{8\times 8}\, \left[  \prod\limits_{i=1}^r \frac{e^{ip_i(n_i-n_{i+1})}}{L^4} 
|\Phi_{n_{i+1}}|^{-2} \DDmtwoRhoInv(p_i)  \left(\hat B^{(p_i)}(\Phi^*_{n_{i+1}})\right) U(p_i,p_{i+1}) \right], \nonumber
\eea
where the definition of the spinor basis transformation matrix $U(p_1,p_2)$ given in Eq.~(\ref{eq:DefOfSpinorBasisTransMat})
was used.

\section{}
\label{AppendixEpsilonCut}
In this appendix we want to deal with the possibly non-positive eigenvalues of the operator $-\kappa^{eff}+\LagrangeMul$,
which would not allow the option of performing the Gauss-integration in Eq.~(\ref{eq:EffectiveActionReducedModel})
over all modes, in a more precise manner. We therefore restart our calculation beginning in Eq.~(\ref{eq:ActionFunctionSigmaModelLambda}).
Now we perform the Gauss-integration solely over those modes $k\in\ImpSpace,\, 0\neq k\neq\pi$ which
have their corresponding eigenvalue of the operator $-\kappa^{eff}+\LagrangeMul$ not smaller than $2\epsilon>0$.
We denote the subset of these modes as $\ImpSpace(\epsilon,\LagrangeMul)$. According to Eq.~(\ref{eq:EigenValuesOfCouplingMatrixP})
it is given as
\beq
\label{eq:DefOfReducedImpSpace}
\ImpSpace(\epsilon,\LagrangeMul) = \Big\{k\in\ImpSpace\,:\,   -\tilde\kappa_N \varphi^2 \sum\limits_{\mu= 1}^{4}\cos(k_\mu) 
- \frac{8\rho^2}{\tilde y_N^2 \varphi^2}\cdot q(k) + \frac{\LagrangeMul}{2}   \ge \epsilon  \Big\}.
\eeq
Performing the Gauss-integration only over these modes the action reduces to
\bea
\label{eq:EffectiveActionReducedModel2}
S[m^i,s^i,\LagrangeMul,\sigma_k^i] &=& -\ln\left[ \mbox{det}''\left( -\kappa^{eff} + \LagrangeMul   \right)   \right]^{-N/2} 
+ \frac{1}{t_N} \Bigg\{ \sum\limits_{i=1}^N \left[m^i\right]^2 \cdot \left\langle0\left| -\kappa^{eff} + \LagrangeMul
\right|0\right\rangle -L^4 \LagrangeMul  \nonumber \\
&+& \sum\limits_{i=1}^N \left[s^i\right]^2 \cdot \left\langle\pi\left| -\kappa^{eff} + \LagrangeMul \right|\pi\right\rangle
+\sum\limits_{i=1}^N \sum\limits_{{k\in\bar\ImpSpace(\epsilon,\LagrangeMul)}\atop{0\neq k \neq\pi}}  \left[\sigma_k^i\right]^2 
\cdot \left\langle k\left| -\kappa^{eff} + \LagrangeMul \right|k\right\rangle
\Bigg\} \\
&=& \frac{N}{2}\mbox{Tr}''\,\ln\left[ -\kappa^{eff} + \LagrangeMul \right] 
+ \frac{N}{\tilde t_N}  \cdot m^2 \cdot L^4 \cdot \left( -8\tilde\kappa_N\varphi^2 - \frac{16\rho^2}{\tilde y_N^2\varphi^2}q(0) + \LagrangeMul  \right) \nonumber \\
&+& \frac{N}{\tilde t_N}  \cdot s^2 \cdot L^4 \cdot \left( +8\tilde\kappa_N\varphi^2 - \frac{16\rho^2}{\tilde y_N^2\varphi^2}q(\pi) + \LagrangeMul  \right)
-\frac{N}{\tilde t_N}L^4 \LagrangeMul \nonumber\\
&+& \sum\limits_{{k\in\bar\ImpSpace(\epsilon,\LagrangeMul)}\atop{0\neq k \neq\pi}} 
\frac{N}{\tilde t_N}  \cdot \sigma_k^2 \cdot L^4 \cdot \left( -2\tilde\kappa_N \varphi^2 \sum\limits_{\mu= 1}^{4}\cos(k_\mu)  
- \frac{16\rho^2}{\tilde y_N^2\varphi^2}q(k) + \LagrangeMul  \right),
\eea
where $\sigma_k^i$ denote the amplitudes of the excluded modes with $k\in\bar\ImpSpace(\epsilon,\LagrangeMul),\, 0\neq k \neq \pi$
and $\bar\ImpSpace(\epsilon,\LagrangeMul)\equiv\ImpSpace/\ImpSpace(\epsilon,\LagrangeMul)$ is the
complement of $\ImpSpace(\epsilon,\LagrangeMul)$.
Here the notation
\beq
\label{eq:RelOfMagToMode2}
\sigma_k^i = \sqrt{\frac{L^4}{N}}\, \sigma_k 
\eeq
was introduced correspondingly to Eq.~(\ref{eq:RelOfMagToMode}) and the plane wave modes $|k\rangle$ were explicitly 
given in Eq.~(\ref{eq:DefOfBRACKETmodes}). The determinant $\mbox{det}''$ and the trace
$\mbox{Tr}''$, respectively, are now only performed over the modes $k\in\ImpSpace(\epsilon,\LagrangeMul),\,0\neq k\neq\pi$,
as desired. The resulting gap equations can now be obtained by differentiating the effective action with respect to
$m,s,\LagrangeMul$ and all $\sigma_k$. This leads again to Eq.~(\ref{eq:LargeYGap1}) and Eq.~(\ref{eq:LargeYGap2}). Only the
third one, Eq.~(\ref{eq:LargeYGap3}), is modified yielding now
\beq
\label{eq:LargeYGapAPP3}
m^2+s^2  + \sum\limits_{{k\in\bar\ImpSpace(\epsilon,\LagrangeMul)}\atop{0\neq k \neq\pi}} \sigma_k^2 \;=  \;
1 - \frac{\tilde t_N}{4} \frac{1}{L^4}\sum\limits_{{k\in\ImpSpace(\epsilon,\LagrangeMul)}\atop{0\neq k \neq \pi}} 
\left[-\tilde\kappa_N \varphi^2 \sum\limits_{\mu= 1}^{ 4}\cos(k_\mu) - \frac{8\rho^2}{\tilde y_N^2 \varphi^2} q(k) +
\frac{\LagrangeMul}{2}\right]^{-1}.
\eeq
Furthermore, one obtains one additional gap equation for every mode $k\in\bar\ImpSpace(\epsilon,\LagrangeMul),\,0\neq k\neq\pi$ 
according to
\beq
\label{eq:LargeYGapAPP1}
0=\sigma_k\cdot \left[\LagrangeMul - \left(  +2\tilde\kappa_N \varphi^2 \sum\limits_{\mu= 1}^{4}\cos(k_\mu)  
+ \frac{16\rho^2}{\tilde y_N^2\varphi^2}q(k)   \right) \right]\quad
\forall k\in\bar\ImpSpace(\epsilon,\LagrangeMul),\,0\neq k\neq\pi .
\eeq
Again we consider the scenario of a purely ferromagnetic phase and the scenario of a purely anti-ferromagnetic phase
for the investigation of the phase structure. The only particularity here is that we assume all $\sigma_k$ to be zero
in both cases. (In principle, with this approach one could also study the phase structure of some of the amplitudes 
$\sigma_k$, but this is beyond our interest here.) We thus arrive directly at the prior equations~(\ref{eq:LargeYGap4}) 
and ~(\ref{eq:LargeYGap6}), respectively, fixing the value of $\LagrangeMul$ as before. With this fixation of $\LagrangeMul$
the subset $\ImpSpace(\epsilon,\LagrangeMul)$ now becomes $\ImpSpace_m(\epsilon)$ for the ferromagnetic phase as already 
defined in Eq.~(\ref{eq:DefOfRedImpSpaceMS}). For the anti-ferromagnetic phase it becomes $\ImpSpace_s(\epsilon)$.
We have now arrived at the final results for the self-consistency equations that were already presented in 
Eq.~(\ref{eq:LargeYGap5}) and Eq.~(\ref{eq:LargeYGap7}). 

In order to get a rough estimate about the validity of neglecting the modes $k\in\bar\ImpSpace(\epsilon,\LagrangeMul)$
one should check the volume of this subset and compare it to the volume of the full set $\ImpSpace$ as we did in
our discussion in the main text.

\end{appendix}

\section*{Acknowledgments}
We thank the "Deutsche Telekom Stiftung" for supporting this study by providing a Ph.D. scholarship for
P.G. We further acknowledge the support of the DFG through the DFG-project {\it Mu932/4-1}.
Furthermore, we are grateful to Michael M\"uller-Preussker and Gian Carlo Rossi for helpful
discussions and comments.

\bibliographystyle{unsrtOWN}
\bibliography{HiggsYukawaPhaseDiagramAnalytic}

\begin{thebibliography}{10}

\bibitem{Smit:1989tz}
J.~Smit.
\newblock Standard model and chiral gauge theories on the lattice.
\newblock {\em Nucl. Phys. Proc. Suppl.} 17:3--16, 1990.

\bibitem{Shigemitsu:1991tc}
J.~Shigemitsu.
\newblock Higgs-{Y}ukawa chiral models.
\newblock {\em Nucl. Phys. Proc. Suppl.} 20:515--527, 1991.

\bibitem{Golterman:1990nx}
M.~F.~L. Golterman.
\newblock Lattice chiral gauge theories: {R}esults and problems.
\newblock {\em Nucl. Phys. Proc. Suppl.} 20:528--541, 1991.

\bibitem{book:Montvay}
I.~Montvay and G.~M{\"u}nster.
\newblock {\em {Q}uantum {F}ields on a {L}attice ({C}ambridge {M}onographs on
  {M}athematical {P}hysics)}.
\newblock {C}ambridge U{}niversity {P}ress, 1997.

\bibitem{book:Jersak}
A.~K. De and J.~Jers{\'a}k.
\newblock {\em {Y}ukawa models on the lattice}.
\newblock {HLRZ} {J\"u}lich, {HLRZ} 91-83, preprint edition, 1991.

\bibitem{Golterman:1992ye}
M.~F.~L. Golterman, D.~N. Petcher, and E.~Rivas.
\newblock On the {E}ichten-{P}reskill proposal for lattice chiral gauge
  theories.
\newblock {\em Nucl. Phys. Proc. Suppl.} 29BC:193--199, 1992.

\bibitem{Jansen:1994ym}
K.~Jansen.
\newblock Domain wall fermions and chiral gauge theories.
\newblock {\em Phys. Rept.} 273:1--54, 1996.

\bibitem{Luscher:1998pq}
M.~L{\"u}scher.
\newblock Exact chiral symmetry on the lattice and the {G}insparg- {W}ilson
  relation.
\newblock {\em Phys. Lett.} B428:342--345, 1998.

\bibitem{Neuberger:1998wv}
H.~Neuberger.
\newblock More about exactly massless quarks on the lattice.
\newblock {\em Phys. Lett.} B427:353--355, 1998.

\bibitem{Nielsen:1980rz}
H.~B. Nielsen and M.~Ninomiya.
\newblock Absence of {N}eutrinos on a {L}attice. 1. {P}roof by {H}omotopy
  {T}heory.
\newblock {\em Nucl. Phys.} B185:20, Erratum--ibid.B195:541,1982, 1981.

\bibitem{Hasenfratz:1991it}
A.~Hasenfratz, P.~Hasenfratz, K.~Jansen, J.~Kuti, and Y.~Shen.
\newblock The {E}quivalence of the top quark condensate and the elementary
  {H}iggs field.
\newblock {\em Nucl. Phys.} B365:79--97, 1991.

\bibitem{Hasenfratz:1992xs}
A.~Hasenfratz, K.~Jansen, and Y.~Shen.
\newblock The {P}hase diagram of a {$U$(1)} {H}iggs-{Y}ukawa model at finite
  lambda.
\newblock {\em Nucl. Phys.} B394:527--540, 1993.

\bibitem{Gerhold:2006rc}
P.~Gerhold and K.~Jansen.
\newblock On the phase structure of a chiral invariant {H}iggs-{Y}ukawa model.
\newblock 2006.

\bibitem{Giedt:2007qg}
J.~Giedt and E.~Poppitz.
\newblock Chiral lattice gauge theories and the strong coupling dynamics of a
  {Y}ukawa-{H}iggs model with {G}insparg-{W}ilson fermions.
\newblock {\em arXiv:} hep-lat/0701004, 2007.

\bibitem{Giedt:2007pri}
J.~Giedt and E.~Poppitz.
\newblock Private communication.
\newblock 2007.

\bibitem{xypd1}
P.~Gerhold and K.~Jansen.
\newblock {\em in preparation}, 2007.

\bibitem{Hernandez:1998et}
P.~Hernandez, K.~Jansen, and M.~L{\"u}scher.
\newblock Locality properties of {N}euberger's lattice {D}irac operator.
\newblock {\em Nucl. Phys.} B552:363--378, 1999.

\bibitem{Flyvbjerg:1988em}
Henrik Flyvbjerg.
\newblock 1/{N} expansion of the nonlinear sigma model.
\newblock {\em Phys. Lett.} B219:323, 1989.

\bibitem{ZinnJustin}
J.~Zinn-Justin.
\newblock {\em {Q}uantum {F}ield {T}heory and {C}ritical {P}henomena}.
\newblock Oxford University Press, 2nd edition, 1993.

\end{thebibliography}

\end{document}